\documentclass{aa}
\usepackage{graphicx,txfonts}
\usepackage[autoalias]{hbnatbib}
\newdimen\captwidth
\newdimen\figwidth
\newcommand{\nai}{\ion{Na}{i}}
\newcommand{\naii}{\ion{Na}{ii}}
\newcommand{\fei}{\ion{Fe}{i}}
\newcommand{\hi}{\ion{H}{i}}
\newcommand{\hei}{\ion{He}{i}}
\newcommand{\hh}{H$_2$}
\newcommand{\caii}{\ion{Ca}{ii}}
\newcommand{\caiii}{\ion{Ca}{iii}}

\newcommand{\feii}{\ion{Fe}{ii}}
\newcommand{\bpw}{\object{$\beta\:$Pictoris}}
\newcommand{\bp}{\object{$\beta\:$Pic}}

\let\dy=\displaystyle
\newcommand{\rd}{\mathrm{d}}
\newcommand{\excs}{\extracolsep{\fill}}
\begin{document}
           %
           %
\title{High latitude gas in the \bpw\ system}
\subtitle{A possible origin related to Falling Evaporating Bodies}
\author{H. Beust \and P. Valiron}
\institute{
Laboratoire d'Astrophysique de Grenoble,
UMR 5571 C.N.R.S.,
Universit\'e J. Fourier, 
B.P. 53, F-38041 Grenoble Cedex 9, France}
\offprints{H. Beust}
\mail{Herve.Beust@obs.ujf-grenoble.fr}
\titlerunning{High latitude gas in the \bpw\ system}
\authorrunning{Beust and Valiron}
\abstract{The puzzling detection of 
\caii\ ions at fairly high latitude ($\ga 30\degr$) above the 
outer parts of the \bpw\ circumstellar disk was recently reported.
Surprisingly,
this detection does not extend to \nai\ atoms, in contradiction with
our modelling of the emission lines in and out of the mid-plane of the disk.} 
{We propose that
the presence of these off-plane \caii\ ions
(and to a lesser extent \fei\ atoms), 
and the non-detection of off-plane \nai\ atoms,
could be the consequence
of the evaporation process of Falling Evaporating Bodies (FEBs),
i.e., star-grazing planetesimals that evaporate in the immediate
vicinity of the star.}
{Our model is two-fold. Firstly, we show numerically and theoretically
that in the star-grazing regime,
the FEBs are subject to inclination oscillations up to
$30$ -- $40\degr$, and that most metallic species released during
each FEB sublimation keep track of
their initial orbital inclination while starting a free expansion 
away from the star,
blown out by a strong radiation pressure. Secondly, 
the off-plane \caii\ and \fei\ species must be stopped prior to their 
detection at rest with respect to the star, about 100\,AU away. 
We revisit the role of energetic collisional processes, and we
investigate the possible influence of magnetic interactions.} 
{This dynamical process of inclination oscillations
explains the presence
of off-plane \caii\ (and \fei). It also accounts for the absence of \nai\ 
because once released by the FEBs, these atoms are quickly photoionized
and no longer undergo any significant radiation pressure.
Our numerical simulations demonstrate 
that the deceleration of metallic ions
can be achieved very efficiently if the
ions encounter a dilute neutral gaseous medium. 
The required \hi\ column density is reduced to
$\sim 10^{17}\,\mbox{cm}^{-2}$, one order of magnitude \emph{below}
present detection limits. We also investigate the possibility
that the ions are slowed down magnetically. While the sole
action of a magnetic field of the order
of $1\,\mu$G is not effective,
the combined effect of magnetic and collisional deceleration processes 
lead to an additional
lowering of the required \hi\ column density by one order
of magnitude.}{}
\keywords{Stars: circumstellar matter -- Stars individual: \bp --
Methods: analytical -- Celestial mechanics -- 
Methods: numerical -- Planetary systems:
protoplanetary disks -- Molecular processes -- Magnetic fields}
\maketitle
\section{Introduction}
The dusty and gaseous disk surrounding the young main-sequence star
\bpw\ has been the subject of intense investigation since its discovery
\citep{st84}. The main motivation for these studies is that
this disk constitutes the most convincing example of a probable
young extrasolar planetary system, possibly analogous to the early
solar system. \bp\ is a young star, but it is not a pre-main sequence star.
Its age has been subject to controversy in past years, but successive
determinations based on the kinematics of the socalled \bpw\ moving
group \citep{bar99,zuck01,or04} lead to a most recent estimate of 11.2\,Myr
\citep{or04}. This shows that
its disk should be called a \emph{young planetary} rather than
a \emph{protoplanetary} disk, meaning that planet formation already should
have had enough time to occur. Indeed, although
no direct planet detection has been made so far, several indirect
observational facts suggest that planets are present in the disk.
This mainly concerns asymmetries found in the disk images from
both scattered stellar light by the dust \citep{kal95,hea00}
and thermal emission by the dust \citep{wei03}, that have been  
modelled as resulting from the gravitational perturbations by 
one Jupiter-sized planet \citep{mou97,hea00,aug01}.

The gaseous counterpart of the dust disk was detected in absorption
in the stellar spectrum \citep{hob85}, and has been regularly
observed since that time. Observations of many metallic species such as
\nai, \caii, \feii\ldots have been reported \citep{vid94},
extending more recently to more fragile species like CO \citep{jol98,lec01}.

The gas was first detected at rest with respect to the star, but 
Doppler-shifted, highly time-variable components are regularly
observed in the spectral lines of many elements
\citep{fer87,xxi,pet99}.
These transient spectral events have been successfully modelled
as resulting from the sublimation of numerous star-grazing planetesimals
(several hundred per year) in the immediate vicinity of the star
\citep[see][and refs. therein]{xxv,xxii}. This scenario has been
termed the \emph{Falling Evaporating Bodies} scenario (FEBs).

From a dynamical point of view, the origin of these numerous star-grazers
seem to be related to mean-motion resonances (mainly 4:1 and 3:1) with
a Jovian planet orbiting the star on a moderately eccentric
orbit ($e'\simeq0.07$--0.1) \citep{bm96,bm00}. In this
context though, the suspected resonance reservoirs are expected to clear out
very quickly. In \citet{pth01}, it was shown that collisions
among the population of the planetesimals constituting the disk
could help replenish the resonances from adjacent regions and
subsequently sustain the FEB activity. However, there are still unsolved
questions concerning this scenario. The main one concerns the amount
of material available \citep{pth03}. There is a large discrepancy
between the number of planetesimals deduced from the FEB model and
that deduced from an extrapolation of the dust observed population
up to kilometre-sized bodies. We nevertheless note in \citet{pth03}
that the mass determination of \citet{pth01} from the FEB scenario
is very imprecise as it is indirect. Indeed if the collisions 
are supplemented by some more violent transport processes,  
then the planetesimals population required to sustain the
FEB activity could be much lower.

An important outcome of this model is that it implies an important
reservoir of planetesimals in the disk and the presence of at least
one giant planet at $\sim 10\,$AU from the star. This is another
argument in favour of the presence of planets in the \bp\ disk.
Moreover, the presence of numerous planetesimals is also a
requirement of the dust models. Due to an intense radiation pressure,
many dust particles should be quickly removed from the system.
The particles observed consist of second
generation material continuously replenished from inside the disk
by planetesimals, either by slow
evaporation \citep{lec96} or by collisions \citep{arty97}.
This justifies the name \emph{second generation} or \emph{debris}
disks given to the \bp\ disk and other similar disks, such as
the HD\,141569 disk \citep[see][and refs. therein]{aug04}.

Radiation pressure affects not only dust particles, but also the
metallic species seen in absorption with respect to the star.  Many of
them undergo a radiation force from the star that largely overcomes
the stellar gravity \citep{xxiv}. This is for instance the case of
\caii\ for which the radiation pressure is 35 times larger than the
stellar gravity. This seems in contradiction with the detection of
circumstellar gas at rest with respect to the star.  \citet{xxiv}
suggested that this stable gas was produced from inside by the FEBs
themselves, that it is then blown away by the radiation pressure, and
afterwards slowed down by a dense enough \hi\ ring where it
accumulates.  The exact shape of this ring is of little importance,
the main parameter being the integrated \hi\ column density, of the
order of $10^{18}\,\mbox{cm}^{-2}$. Detailed modelling shows that all
stable circumstellar lines can be reproduced this way.

In a recent paper, \citet{bra04} report the detection with VLT/UVES of
\emph{emission} lines of metals (\fei, \nai, \caii) in the \bp\ disk,
i.e., away from the direction of the star. They report the detection
of \nai\ and \fei\ up to more than 300\,AU from the star.  \nai\ was
resolved earlier by \citet{olof01}, but the detection
of \citet{bra04} extends further out. $\mathrm{H}_2$ was also claimed
to be detected in emission by \citet{thi01}, implying huge quantities
($\sim 50\,M_\oplus$) in the \bp\ system, but this was questioned by
\citet{lec01}, who reported from FUSE/LYMAN observations an upper
limit $N(\mathrm{H}_2)\la10^{18}\,\mathrm{cm}^{-2}$ for the
$\mathrm{H}_2$ column density towards \bp.

A particularly puzzling outcome of the \citet{bra04} observations
is the detection of \caii\ emission at fairly high latitude above the
mid-plane of the disk. \caii\ is detected at 77\,AU height above and below
the mid-plane at 116\,AU from the star. This corresponds to an inclination
of $33\degr$ above the mid-plane. At this distance, \caii\ is detected
in both branches of the disk and on both sides of the mid-plane, and
the emission at $33\degr$ inclination largely overcomes that in
the mid-plane. Surprisingly, the \fei\ and \nai\ emission do not
exhibit such a structuring. It is conversely concentrated in the mid-plane
of the disk. However, the \fei\ emission is broader than the \nai.
At the height above the mid-plane corresponding to the peak emission
in the \caii\ lines, the \fei\ is still detected.
The gas shares the same radial velocity as the star
within 1 or 2\,km\,s$^{-1}$ at most.

\emph{There is no straightforward explanation for the presence of
species like \caii\ at such latitudes above the disk, nor for the
absence of other species.}  The purpose of this paper is to propose
that this gas could constitute material released by the FEBs in the
vicinity of the star, first blown away by radiation pressure, and then
stopped far away from the star by friction with some gaseous medium,
and/or by magnetic interaction.  The \caii\ and possibly the \fei\
reach a significant inclination because the parent bodies (the FEBs)
initially orbiting within the plane of the disk undergo inclination
oscillations up so several tens of degrees when they reach the
star-grazer state. Once released by the FEBs, the ions keep track of
that inclination. In Sect.~2, we model the formation of the emission
lines in and out of the mid-plane of the disk. We show that all the
emissions are compatible with solar relative abundances between the
elements under consideration, except that sodium should necessarily
not be present (or strongly depleted) in the high latitude gas.  In
Sect.~3, we expose the dynamical model for high latitude gas
generation, and we detail the theoretical background for inclination
oscillations in the FEB state. We show that due to a negligible
radiation pressure on \naii, sodium should not be present in this gas,
in agreement with the observations.

In Sect.~4, we investigate how the \caii\ ions could be slowed down at
the stellar distance they are observed.  We show that this
deceleration can be achieved by collision with a dilute neutral
medium, and we discuss the role of elastic and inelastic collisional
processes.  We also discuss the effect of a non-radial magnetic
field. While the magnetic field in itself is inefficient to decelerate
the ions, we show that its presence increases the efficiency of the
deceleration by a neutral medium by an order of magnitude or more.
\section{Modelling the emission lines}
\begin{table}
\caption[]{Measured circumstellar ratios, and simulated emission
intensities for the transitions under consideration. The circumstellar
factors are taken from calibrated ESO/HARPS spectra of \bp\
(Galland F., private communication). The emission intensities (given
in arbitrary units) out of the mid-plane are derived from the
\textsc{Cloudy} run and those in the midplane are estimated by
multiplying the former ones by the circumstellar factors.}
\label{fik}
\begin{tabular*}{\columnwidth}{@{\excs}llll}
\hline\noalign{\smallskip}
Transition & Circumstellar & Emission           & Emission\\
           & factor        & intensity          & intensity\\
           &               & above the mid-plane & in the mid-plane\\
\noalign{\smallskip}\hline\noalign{\smallskip}
\nai\ D$_1$     & 0.91   & 12.5 & 11.4  \\
\nai\ D$_2$     & 0.90   & 24.8 & 22.3  \\
\caii\ K        & 0.033  & 15.8 & 0.521 \\
\caii\ H        & 0.0057 & 15.6 & 0.0889 \\
\fei\ 3859\,\AA & 0.86   & 23.9 & 20.55  \\
\noalign{\smallskip}\hline
\end{tabular*}
\end{table}

Our first task is to model the emission lines of \caii, \nai\ and \fei\
observed in and out of the mid-plane of the disk, as observed by
\citet{bra04}. We focus on the \caii~K and H lines, the \nai\ D$_1$ and
D$_2$ lines, and the \fei\ $\lambda=3859\,$\AA\ line. The modelling is
done using the radiative transfer code \textsc{Cloudy} by \citet{fer98}.

We take the synthetic \textsc{Atlas9} \citep{kur} stellar model for
\bp\ with $T_\mathrm{eff}=8100\,$K, $\log g=4.207$ and
a total luminosity of $8.7\,\mathrm{L}_\odot$. We put gas at 116\,AU
from the star and compute the line emission. The chemical composition
of the gas is assumed to be solar. The gas is modelled as a
10\,AU wide layer with solar composition and a given hydrogen density.

Without any additional energy source other than the stellar radiation
flux, the emission in all lines as computed by \textsc{Cloudy} appears
negligible, because the gas remains very cold (a few Kelvins). Thus,
in order to generate detectable lines, the gas must be heated by some
energy source.  There can be some turbulence, but in the framework of
our model, the strong deceleration of the weak flux of incoming
metallic ions might constitute a sufficient heating source.  In the
following we thus assume the emission lines to be excited thermally.
The incoming ions, once blown away by the radiation pressure, reach a
velocity of $\sim 10^3\mbox{km}\,\mbox{s}^{-1}$ at 100\,AU. The
various surveys of the FEB activity towards \bp\ \citep{xxii} led to
estimate that one roughly 1 kilometer-sized body is destroyed in front
of the line of sight every day, with temporal fluctuations around one
order of magnitude. Assuming that not all FEBs cross the line of
sight, we estimate the total number of FEBs evaporated as $N=\sim 10$
per day. Taking kilometer-sized bodies with an average density of
$2\,\mbox{g}\,\mbox{cm}^{-3}$, assuming that half of their mass is
made of metallic ions that are pushed away by radiation pressure, and
assuming that the ions spread over an open cone of $\alpha=\sim
30\degr$ half-opening angle (in order to disperse ions at that
latitude), we may estimate the incoming kinetic energy flux $F$ at
100\,AU as
\begin{equation}
F=\frac{1}{2}\frac{Nmv^2}{4\pi d^2\sin\alpha}
\sim 0.2\,\mbox{erg}\,\mbox{cm}^{-2}\,\mbox{s}^{-1}\quad,
\end{equation}
where $v\simeq1000\,\mbox{km}\,\mbox{s}^{-1}$ is the velocity of the
incoming ions and $m$ is the average mass of the FEBs.
This kinetic energy heats the
local \hi\ gas over a distance corresponding to the stopping path of
the metallic ions. This depends on the density of the local gas, but
we see in Sect.~4  that an \hi\ column density of
$\sim 10^{17}\,\mbox{cm}^{-2}$ is expected to be sufficient to stop
the ions over a distance of a few AU.  

We thus decided to perform
runs of \textsc{Cloudy} with an extra heating source, with a volume-heating
rate (parameter \texttt{hextra}) corresponding to the incoming kinetic
energy flux deposited over the stopping distance.
We performed several runs for 
hydrogen densities ranging
between $10^6$ and $10^{10}\,\mbox{cm}^{-3}$. The result listed in
Table~\ref{fik} are for $10^7\,\mbox{cm}^{-3}$. For other values, the 
absolute values of the emissions changes, but their relative
behaviour remains comparable.

In almost all runs the same line behaviour is reported
(Table~\ref{fik}): The \caii\ K and H emissions are comparable, the
\caii\ K emission being slightly stronger; the \nai\ D$_2$ emission is
typically twice as strong as the \caii\ K one, while the \feii\
emission is as strong as the \nai\ one. There is only little variation
between the ``hot'' and ``cool'' cases.

This simulation is intended to hold for the gas \emph{out} of the
disk mid-plane. The spectrum of \bp\ as seen from Earth is known to
present stable circumstellar components \citep{xxiv} due to the
gaseous counterpart of the disk. These componenents appear as sharp
($\sim 0.1\,$\AA\ wide) additional absorptions at the bottom
of the rotationally broadened photospheric stellar lines.

There is a major difference
between ions located in and out of the mid-plane of the disk~: the
former ones see a stellar spectrum \emph{with} these circumstellar
components, while the latter ones see a stellar spectrum
\emph{without} these components (because they view the star across
the disk, as Earth observers).
The \textsc{Atlas9} stellar model does not take into account these
circumstellar absorptions. Hence we must add them to the model.
Unfortunately, adding additional
spectral absorptions to the stellar models provided
is not a standard procedure for \textsc{Cloudy}.
It is thus not possible to accurately model the emission in the
mid-plane of the disk. A first order approximation in order to take
these absorptions into account is to apply 
reduction factors of the stellar flux to ions
in the mid-plane of the disk (and only to them).
These factors are defined as the
ratio of the stellar flux at the bottom of the circumstellar
additional absorptions (if present) to the flux at the bottom of
the corresponding photospheric lines
(or equivalently the top of the circumstellar lines). They are simply
measured from observed \bp\ spectra. We used ESO/HARPS spectra
communicated by F. Galland.
The factors are listed in Table~\ref{fik}. We note that the circumstellar
lines are particularly deep for the \caii\ lines.

In order to derive a rough estimate of the line emissions,
we may assume that the 
emission is proportional to the incoming flux. This only
applies if the lines are optically thin, but the ratio of 2
between the \nai\ D$_2$ and D$_1$ emission shows that this is the case
here.
We apply the
circumstellar factors of Table~\ref{fik} to the emission intensities
derived from \textsc{Cloudy} with no circumstellar absorptions.
The ratio between
\nai\ and \fei\ emissions remains unchanged (thanks to similar
circumstellar factors), but the \caii\ emissions (in both lines)
appears now far weaker (a few hundredth in relative intensity) 
than the \nai\ and \fei\ ones.  

The observational constrains to fulfill \citep{bra04}
are the following: In the mid-plane, the \caii\ emission is small
compared to the \nai\ and \fei\ ones; the \nai~D$_2$ to \nai~D$_1$
ratio is close to 2, showing that the emission is optically thin;
the emission in the \fei\ $\lambda=3859\,$\AA\ line is comparable to
that in the \nai~D$_2$ line. Out of the mid-plane, the \caii\ emission
dominates, but the \fei\ line may still be detected, because the
wing of the line  is much larger than for the \nai\ lines; the
\caii\ K to \caii\ H ratio is close to one.

Our simple model succeeds in reproducing the emissions in
the mid-plane. Out of the mid-plane though, the \nai\ emission is
always stronger than the \caii\ one and comparable to the
\fei\ one. It is thus impossible to simultaneously have a strong
\caii\ emission, a \fei\ emission a few times weaker, and an
undetectable \nai\ emission. The only possibility is to exclude the
hypothesis of solar composition.

We come therefore to the following conclusions~: 1) for 
the emission in the mid-plane, the observations can be reproduced
assuming solar relative abundances between iron, sodium and calcium;
2) out of the mid-plane, the \fei\ and \caii\ line intensities can be
consistently simulated assuming solar relative abundances; 3)
the non-detection of \nai\ emission out of the mid-plane cannot be
explained in these conditions, unless sodium is strongly depleted with
respect to solar abundance. The model presented in next Section
provides a plausible explanation for such sodium depletion.
\section{Falling Evaporating Bodies and high latitude ions}
\subsection{General features}
We propose an origin for the high latitude ions observed
by \citet{bra04}. Two facts need to be explained: i) Why are there
large amounts of gas at $30\degr$ above the mid-plane of the disk ?
ii) Why does sodium seem not to be present in this gas ? We propose that
this high latitude gas could be produced by the Falling Evaporating
Bodies (FEBs).

The FEBs are star-grazing bodies that fully evaporate in the vicinity
of the star. Dynamically speaking, they are planetesimals that have
been extracted from the disk orbiting the star and driven to high
eccentricity orbits. They enter the FEB regime when their periastron
reaches a threshold value ($\sim 0.4\,$AU) that allows the refractory
material to evaporate.  The details of the evaporation process of the
bodies as their periastron decreases down to a few stellar radii are
described in \citet{ckar2}. The bodies start to evaporate at each
periastron passage, and their evaporation rate increases as the
periastron distance gets smaller.  For the sizes considered ($\sim
10\,$km), the FEBs are fully evaporated when they reach a periastron
value $q\simeq0.15\pm0.05\,$AU.

Star-grazers may be produced from a disk of planetesimals
by planetary perturbations. The most efficient mechanism is the Kozai
resonance \citep{koz62}, which concerns bodies that have initial
high inclination with respect to the orbital plane of the planetary
system. Under secular perturbations, the initially highly inclined
body is periodically driven to low inclination, but very eccentric,
star-grazing orbits. This mechanism is responsible for most of the
sun-grazing bodies reported in the Solar System, such as the Kreutz
group \citep{bai92}. 

However, the Kozai resonance is due to the secular, circular part of
the interaction Hamiltonian with the planet(s). It is therefore
invariant with respect to any rotation in the planet's orbital plane.
Bodies driven by the Kozai mechanism are thus expected to reach the
FEB state with random orbit orientations. This does not match the
statistics of the Doppler velocities of the variable spectral events
observed, which shows a strong bias towards redshifts. Most of the
suspected FEBs share some kind of common preferred periastron
orientation range which is not compatible with the Kozai resonance
\citep{xxii}.

\citet{bm96} proposed that the FEBs could be generated by another
mechanism involving mean-motion resonances with at least one major
perturbing planet. The secular motion of bodies trapped in a given
mean-motion resonance with a planet is usually characterised by coupled
oscillations of the semi-major axis and the eccentricity around a
median value, but if the planet's orbit is slightly eccentric,
these oscillations are superimposed on a long-term drift
of the eccentricity that can in some cases bring it to star-grazing
values. \citet{yosh89} showed that these changes are particularly
important for resonances 4:1, 3:1 and 5:2. \citet{bm96} showed that
the 4:1 resonance is a potential source of FEBs via this mechanism,
and in \citet{pth01}, it was shown that the 3:1 may also contribute
to the FEB phenomenon. The planet's eccentricity $e'$ does not need
to be very high; $e'\ga 0.05$ is enough, $e'=0.07$ or 0.1 being typical 
convenient values. Such eccentricity values are regularly reached
by Jupiter due to its secular evolution. This mechanism is
close to the one that gave birth to the Kirkwood gaps in the asteroid
belt, even if in the latter case, the overlapping of mean-motion resonances
with secular resonances considerably enhances the mechanism \citep{mm93,
mm95,mmr95,far94}. We cannot exclude this enhancement as being effective
in the \bp\ system (this would imply the presence of more than one
planet), but there is no way to constrain it. We are perhaps 
witnessing in the \bp\ system a process similar to what occurred
in the Solar System, as it was of comparable age to \bp\'s current age.

Due to this mean-motion resonance mechanism, 
a given body, initially orbiting the star in 3:1
or 4:1 mean-motion resonance with a Jupiter-sized planet, may reach
the FEB state within $\sim 10^4$ orbital periods of the planet. 
Contrary to the Kozai case, 
the periastron longitudes of
the FEBs at high eccentricity are now constrained by that
of the perturbing planet, and share some common orientation 
in closer agreement with the observations.
This scenario was numerically tested over a large number of particles
\citep{bm00,pth01}, using the popular symplectic integration package
SWIFT\_MVS \citep{wh91,ld94}. It was shown that the suspected mechanism
was able to fairly well match the statistics of the observed
FEB velocities, provided the orbit of the perturbing planet adopts
a given longitude of periastron with respect to the line of sight. 
If the disk of planetesimals holds a large enough population of bodies,
collisions may help refill the resonance and sustain the FEB activity
\citep{pth01}.

\begin{figure}
\includegraphics[angle=-90,width=\columnwidth]{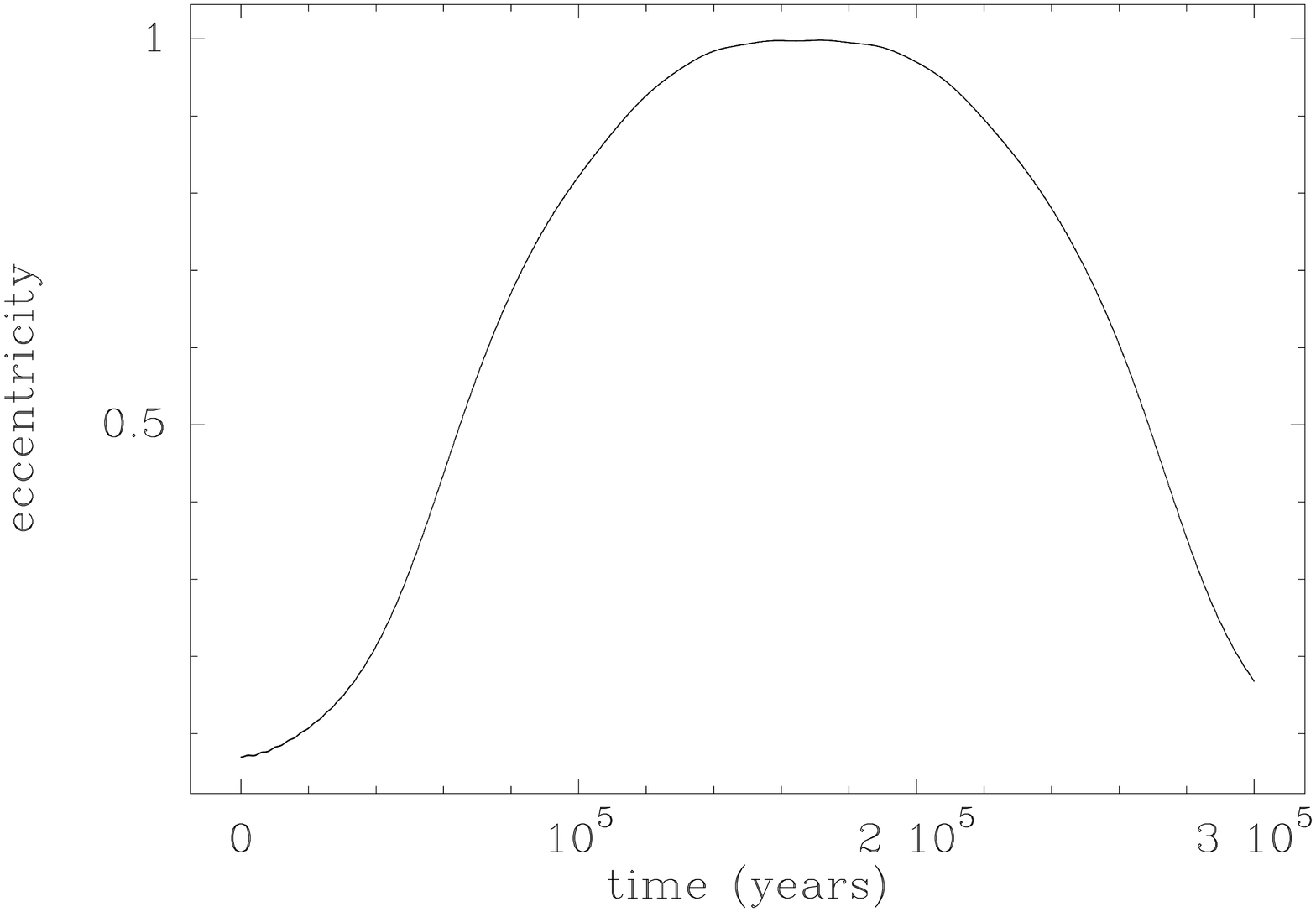}
\includegraphics[angle=-90,width=\columnwidth]{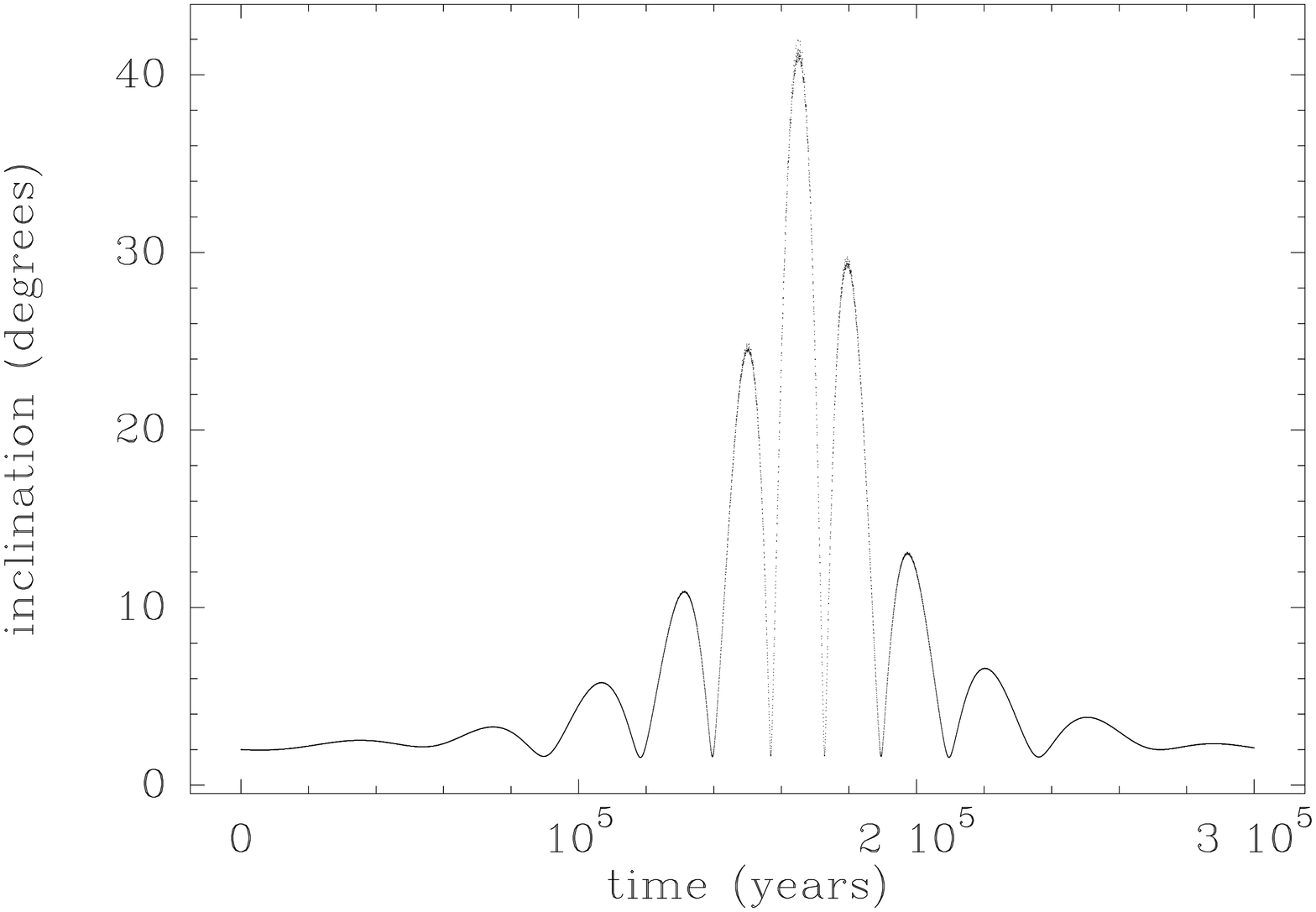}
\caption[]{Temporal evolution of the eccentricity (top) and
of the inclination (bottom) of a typical particle trapped in 4:1 resonance
with a planet orbiting \bp\ at 10\,AU with eccentricity $
e'=0.07$. The planet's mass is 1/1000 of that
of the star. The initial eccentricity of the particle is 0.05 and its
initial inclination is $2\degr$.}
\label{eccinc}
\end{figure}
The orbits of the FEB progenitors in the mean-motion resonances are supposed to
be roughly coplanar with the plane of the disk. In the simulations of
\citet{bm00} and \citet{pth01}, the initial inclinations of the particles
with respect to the orbit of the perturbing planet were initially
chosen as less than $5\degr$ and $3\degr$ respectively, in order to mimic
the typical distribution within a cold planetesimal disk. During their
evolution within the resonance, as long as their eccentricity grows,
the inclination of the particles remains small, but as they reach the FEB
state close to $e\simeq 1$, their inclination is subject to oscillations
of larger amplitude, up to several tens of degrees. This is illustrated in
Fig.~\ref{eccinc}, which shows the secular evolution of the eccentricity
and of the inclination of a typical particle trapped in 4:1 resonance
with a planet orbiting \bp. The particle starts at eccentricity $e=0.05$
and evolves towards the FEB state at $e\simeq 1$ (the peak eccentricity
is about 0.998), and then starts a decrease of its eccentricity. Of course
the decrease phase is purely fictitious, as a real FEB would be destroyed
by the successive periastron passages at peak eccentricity. The
inclination, initially set at $2\degr$, remains small for
$\simeq10^5\,$yr and then starts oscillations that brings it far above
the initial value, up to $40\degr$. This oscillation regime
stops after $\sim 2.5\times10^5\,$yr; it corresponds to the
high eccentricity phase, when $e\ga 0.85$. Hence the FEBs 
may remain within the disk during most of their secular evolution,
but finish in the FEB state with inclinations  that might bring them
significantly out of the disk. The FEB themselves do not
have excursions far out of the plane of the disk, because, as we will
see below, when their inclination is high, their argument of
periastron $\omega$ is close to $0\degr$ or $180\degr$. This causes the
major axis of their orbit to lie roughly in the plane of the disk at the time
the inclination is high. As the orbit is very eccentric, the vertical
excursion of the particle is limited. This is why this effect,
present in the simulations of \citet{bm00} and \citet{pth01},
does not have much influence on the visibility of the FEBs (they need to
cross the line of sight to be detected in absorption).

What is true for the parent body is not necessary true for its byproducts.
The metallic ions released by the FEBs such as \caii\ start to expand
radially around the nucleus and stay for a while in a surrounding
cloud that enables
the FEB to be detected in absorption when it crosses the line of sight;
but they are quickly expelled from there by the intense radiation pressure
they suffer and then start a free expansion out of the system. This process
is extensively described in dedicated simulations in \citet{x,xxii,xxv}.
Dynamically speaking, the ratio of the radiation pressure to stellar
gravity is a constant (usually noted $\beta$) for a given ion or dust
grain. This is equivalent to the view that
with radiation pressure, the ion feels the gravity of star as if its
mass was multiplied by $1-\beta$. For \caii, $\beta>1$
($\beta=35$), so that the ions feel a negative mass and are strongly repelled
by the star. They nevertheless follow a purely Keplerian orbit, namely
a hyperbolic repulsive one, similar to the relative motion of two charged
particles with charges of the same sign. This orbit is very
different to that of the parent body. Both orbits share however the 
\emph{same orbital plane}. Indeed, the ejection velocity of the material
escaped from the FEB \cite[typically $\sim 1\,\mathrm{km}\,\mathrm{s}^{-1}$;
see refs. in][]{xxii} is very small compared to the orbital velocity
of the FEB itself at a few stellar radii from the star (typically
several hundreds of km\,s$^{-1}$). The \caii\ ions may thus be considered
with reasonable accuracy to have the same orbital velocity at ejection
time as their parent body. Both Keplerian orbits are very different
because of radiation pressure, but they share roughly the same orbital
plane. In particular, the \caii\ ions keep memory of the orbital inclination
of their parent body at ejection time, even if this inclination is large
thanks to inclination oscillations in the FEB state. But contrary to the 
parent body, the orbit of \caii\ ions are not confined close to the plane. 
As explained above, the argument of periastron $\omega$
of the parent bodies is close to 0 or 180\degr\ in the high inclination state;
this forces the FEB to remain close to the mid-plane of the disk. 
Due to radiation pressure, the shape of the \caii\ ions is very different from
that of their parent body, and their argument of periastron is not constrained
in the same way. If they have
a high initial inclination, they may evolve far off the plane of the disk
as they escape from the system. If some dense medium is present at a given
distance to brake them, they may be detected as an extended emission
significantly above the plane such as in the \citet{bra04} observation.
Their detection of \caii\ at $33\degr$ off the mid-plane could then
well correspond to ions that have been produced close to the star
by FEBs with similar inclination, and that have freely escaped
up to 100\,AU before being stopped there. 

Then, why should this process only concern \caii\ and not the other
species detected in emission by \citet{bra04} ? Iron and sodium are
byproducts of dust sublimation in FEBs like calcium. Contrary to
\caii, \nai\ and \fei\ are quickly photoionized by the star in the FEB
environment. \feii\ (like \fei) still undergoes a radiation pressure
that overcomes the stellar gravity \citep{xxiv}, so that iron is
expected to behave like calcium.  However, unless the electronic
density is high, as is the case in the vicinity of the mid-plane disk,
iron remains predominantly in the \feii\ state whose spectral lines
were not searched for by \citet{bra04}.  Conversely, \naii\ does not
feel any noticeable radiation pressure, so that once produced, the
\naii\ ions keep following the original orbit of the FEB. They may
afterwards diffuse slowly in the mid-plane of the disk, but they are
not subject to a quick off-plane ejection like the other species. In
this context, we thus expect the gas expelled off-plane by this
process to contain calcium and iron with solar relative abundances,
but no sodium.  This matches our analysis of the emission lines.

In the following we detail the theoretical background of the origin of
the inclination oscillations of the FEBs in high eccentricity regime,
and we show examples from the simulations from \citet{bm00} and
\citet{pth01}.
\subsection{Three-dimensional motion in mean-motion resonance}
The theoretical background for the FEB dynamics in mean-motion resonance
is described in \citet{bm96}, but restricted to the \emph{planar}
case.
Here we wish to extend it to three-dimensional motion. We describe
the restricted three-body problem, i.e., a problem
where a mass-less test particle orbiting a star is perturbed by a
planet orbiting the star on an unperturbed Keplerian orbit.

The full analytical analysis is presented in Appendix~A. We assume
that the particle is locked in a $(p+q):p$ mean-motion resonance with
the planet. The resonant motion is usually described by the
``critical angle of the resonance'' $\sigma$ \citep{mm95}, with
\begin{equation}
\sigma=\frac{p+q}{q}\lambda'-\frac{p}{q}\lambda-\varpi\qquad,
\end{equation}
where $\lambda$ is the mean longitude of the particle
along its orbit; $\lambda'$ is the same for the planet;
$\varpi$ is the longitude of periastron. 

Non-resonant orbits are characterised by a more or less
regular circulation of $\sigma$, while resonant orbits exhibit
libration of $\sigma$ around a stable position. If the planet's orbit
is circular, then the following quantity is a secular constant of
motion \citep{mm93,mm95}~:
\begin{equation}
N=\sqrt{(1-\mu)a}\left(\frac{p+q}{p}-\sqrt{1-e^2}\cos i\right)\qquad,
\end{equation}
where $e$ is the eccentricity, $i$ is the inclination, and $\mu$ is
the mass parameter, i.e. the ratio of the planet's mass to the total
mass. The inclination oscillations are controlled by this
parameter. If the planet's orbit is not circular, then strictly
speaking $N$ is no longer constant, but as the planet's eccentricity
$e'$ is moderate, its variation is slow.

\begin{figure*}
\makebox[\textwidth]{
\includegraphics[width=0.49\textwidth]{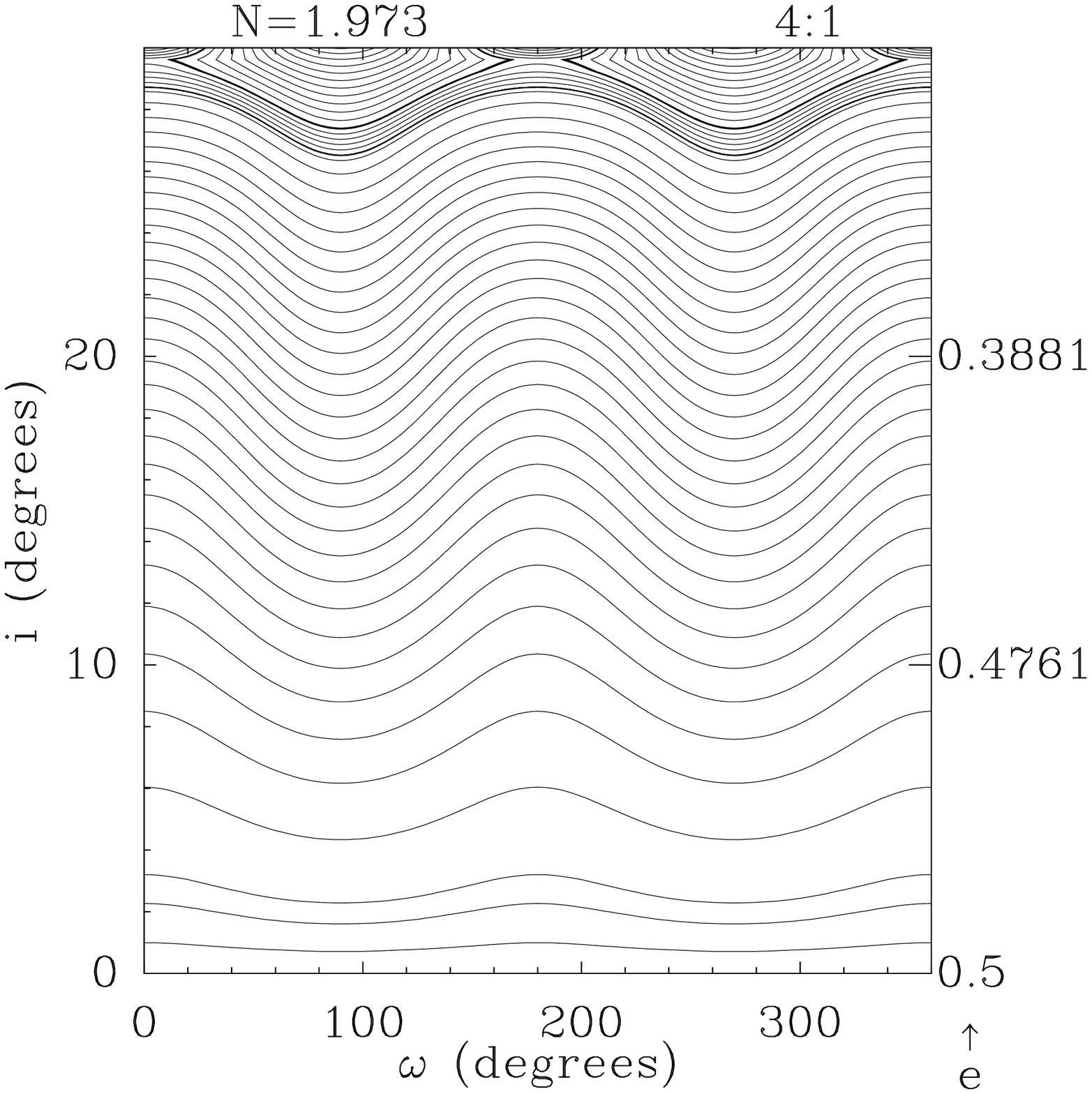}\hfil
\includegraphics[width=0.49\textwidth]{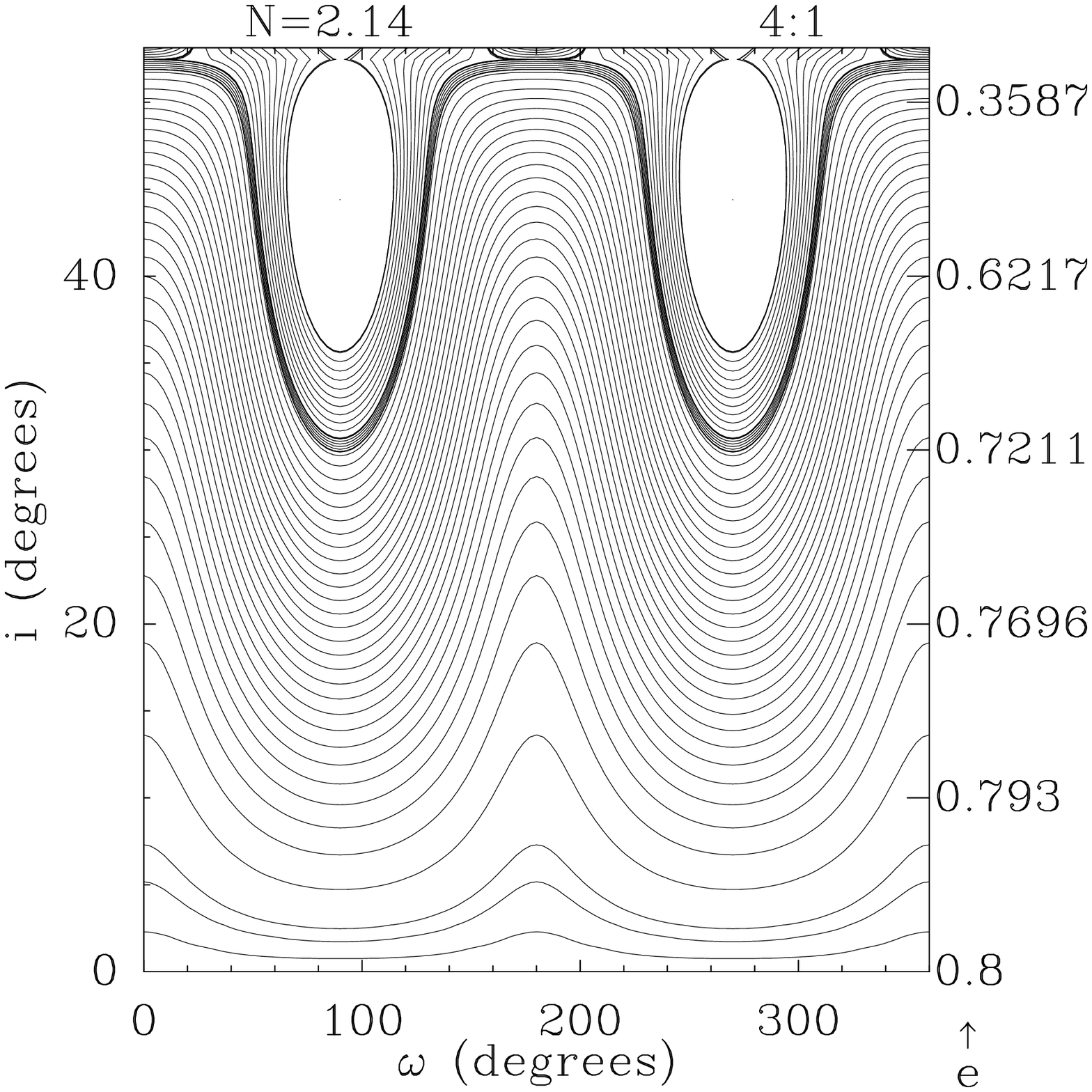}}
\makebox[\textwidth]{
\includegraphics[width=0.49\textwidth]{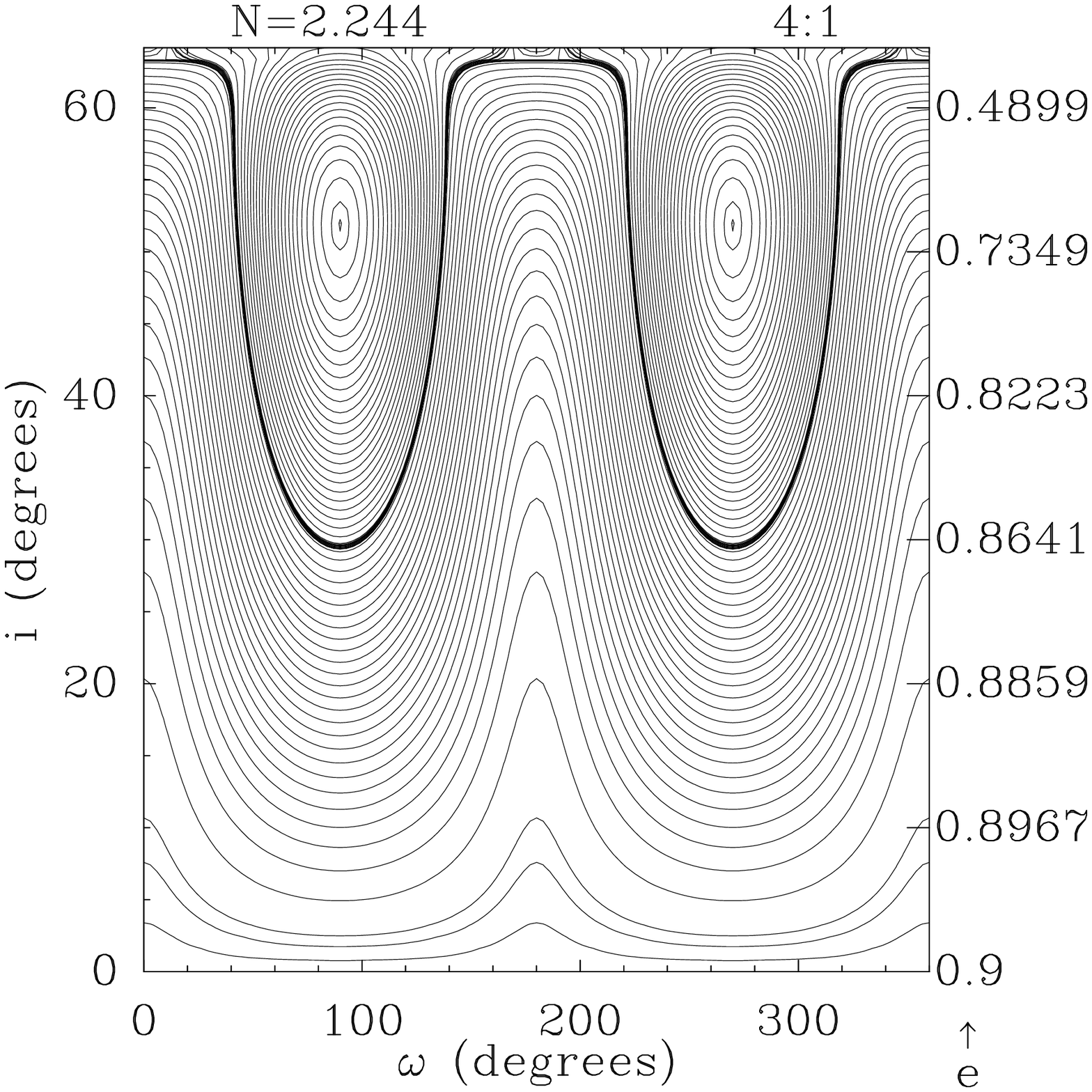}\hfil
\includegraphics[width=0.49\textwidth]{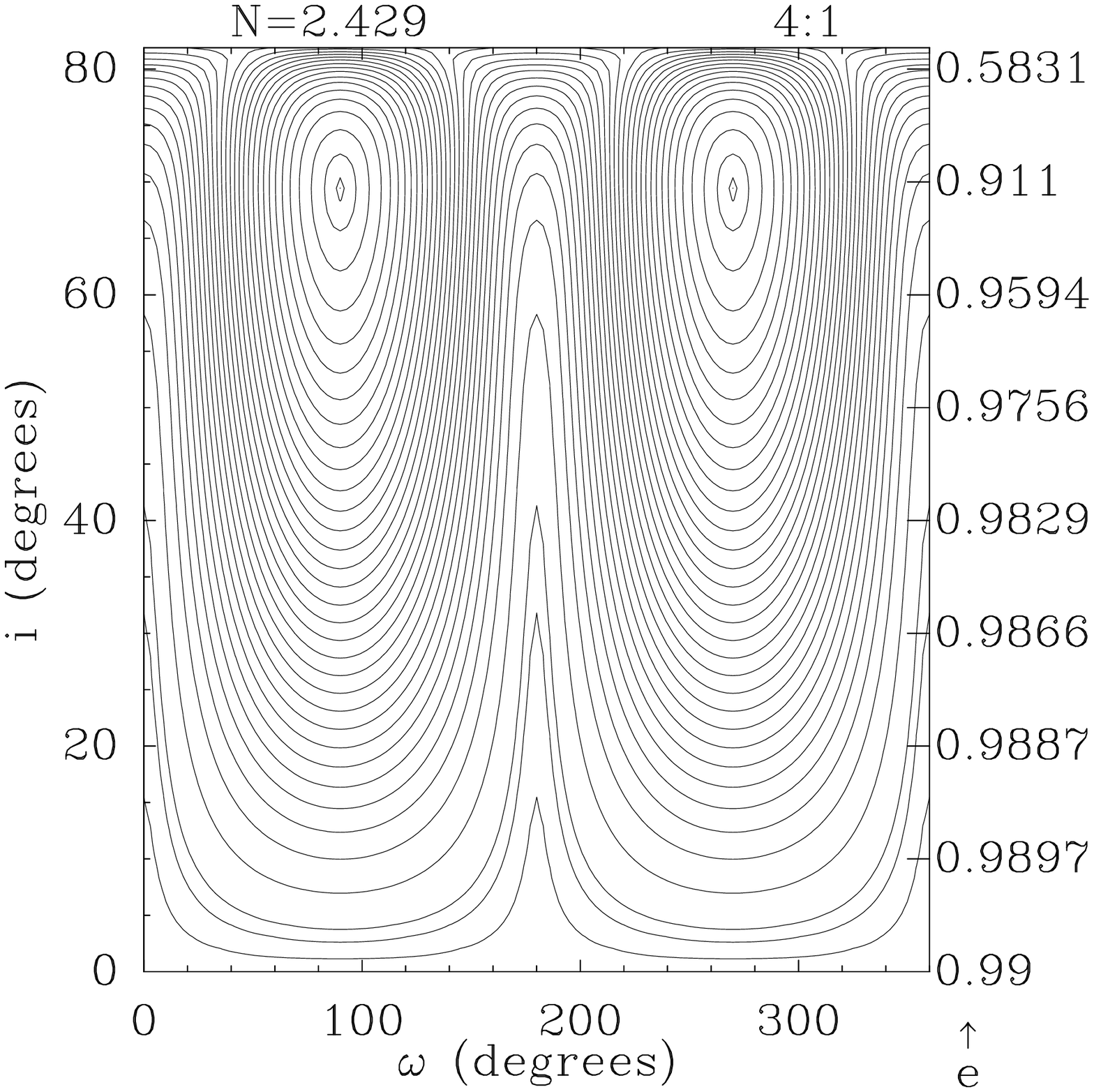}}
\caption[]{Level curves of the Hamiltonian $\mathcal{H}$ in the
$(\omega,i)$ plane for the circular problem, for particles having
negligible $\sigma$-libration amplitude, for different values of the
constant parameter $N$. The eccentricity scale (denoted ``e'' on the
right of the plots) is related to the inclination scale to the left by
the constant value of $N$.}
\label{hseci}
\end{figure*}
Considering the circular problem is equivalent to expanding
the Hamiltonian $\mathcal{H}$ in powers of $e'$,
and retaining only the leading term. The leading
(circular) term is responsible for the $\sigma$-libration, but
also for the inclination oscillations at high eccentricity. The
higher order terms in $e'$ cause a slow drift of the $N$ parameter.
This drift can drive the particle to high eccentricity, even in the
planar problem; this is the way the FEBs are generated.

Finally, the dynamics
of the particle is characterised by three time-scales~: a first,
small one related to the $\sigma$-libration; a second, larger one
characterising the inclination oscillations; a third, long one
describing the secular eccentricity changes from $\sim 0$
to $\sim 1$. The second time-scale is larger than the first
one, but significantly smaller than the third one. Hence during
one inclination oscillation, the value of $N$ may be considered
as $\sim$\,constant, which is equivalent to considering the circular 
problem. 

In the circular problem, the secular Hamiltonian $\mathcal{H}$
depends only on the inclination $i$ and on the argument of periastron
$\omega$, once the value of $N$ is fixed.
It is possible
to explore the dynamics just drawing level curves of  $\mathcal{H}$ 
in $(\omega,i)$ plane, exactly as done in \citet{bm96}. This
is done in Fig.~\ref{hseci} for the 4:1 resonance, for 4 different
values of $N$. 
Instead of giving the value of $N$, we give a value
for the eccentricity and compute the value of $N$ that gives
this eccentricity value for $i=0$.
The value of $\mu$ was fixed to 0.001 as typical for a Jupiter-sized
planet.  The value of $N$ is indicated above each plot, and
a corresponding eccentricity scale is given to the right of the
plots. The four plots corresponds to eccentricity values at $i=0$ of
0.5, 0.8, 0.9 and 0.99 (these values appear at the lower right
corner of the plots). In all these plots,
the eccentricity variation over most of the plot is very moderate.
Hence each of these plots should be regarded as a picture of
the dynamics in a given eccentricity regime. The plots of Fig.~\ref{hseci}
are equivalent to those of Fig.~5 from \citet{mm93} for the 2:1
resonance.

Consider now a given particle trapped in the 4:1 resonance which
starts its eccentricity growth. In a low eccentricity regime,
like the one for $e=0.5$ ($N=1.973$), the inclination remains
low while $\omega$ circulates. Following a level curve of $\mathcal{H}$,
if the inclination is initially low (a few degrees), it undergoes
small variations that keep it in the same range. At a higher eccentricity
regime, the phase portrait changes. We note in Fig.~\ref{hseci}
that two islands of libration for $\omega$ appear around $\omega=\pm 90\degr$.
However, these islands of libration do not concern the particles
we are considering. Our particles start within the plane of the disk
with an inclination that does not exceed a few degrees. Hence the curves
they follow are those located \emph{below} the islands of libration.
For our particles, $\omega$ still circulates, but 
following the level curves, the inclination $i$ is subject
to periodic jumps up to possibly several tens of degrees when $\omega$ reaches
0 or $\pi$. The higher the eccentricity regime, the higher the
inclination jumps. This is the origin of the inclination
oscillations reported in the numerical integration. 

This dynamics is a resonant version of the Kozai dynamics. In the
non-resonant circular restricted problem, the Kozai Hamiltonian
describes the secular dynamics of the particle. It is obtained by a
double averaging of the original Hamiltonian over the orbital motions
of the planet and of the particles \citep{kin99}. It is well known
that this Hamiltonian has a secular constant of motion which is the
$z$-component of the angular momentum (or equivalently
$\sqrt{1-e^2}\cos i=\mathrm{cst}$). It is also well known that at high
inclination, this Hamiltonian shows two islands of libration in
$(i,\omega)$ space around $\omega=\pm\pi/2$, and that particles moving
in these islands evolve periodically from a high inclination and low
eccentricity regime to a low inclination and high eccentricity. This
behaviour constitutes the Kozai resonance \citep{koz62}.

The islands of libration in the plots of Fig.~\ref{hseci}
describe a Kozai resonance, within a mean-motion resonance. 
Indeed, as $a$ is fixed the condition $N=\mathrm{cst}$
is exactly equivalent to the Kozai condition 
$\sqrt{1-e^2}\cos i=\mathrm{cst}$. The FEBs trapped in the 4:1 resonance
that evolve at very high eccentricity regimes are concerned by
this, but they are \emph{not} trapped into the Kozai resonance,
as their argument of periastron $\omega$ still circulates, and 
as they periodically return to $i\simeq 0$. However the Kozai
dynamics influences them and causes periodic inclination jumps
up to several tens of degrees, even if the initial inclination
is low (a few degrees).
\subsection{Tests over a large number of bodies}
\begin{figure}
\includegraphics[angle=-90,width=\columnwidth]{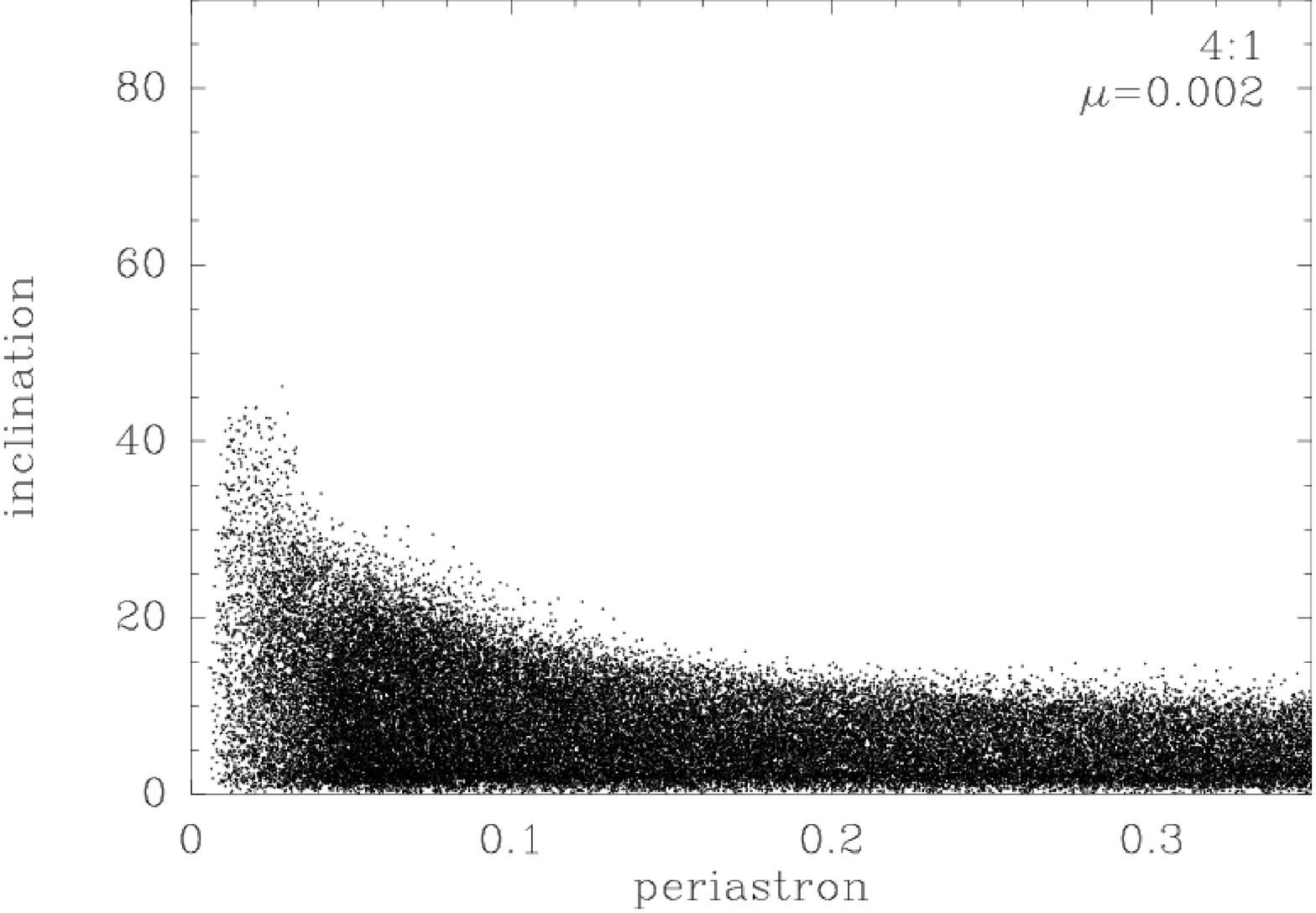}
\includegraphics[angle=-90,width=\columnwidth]{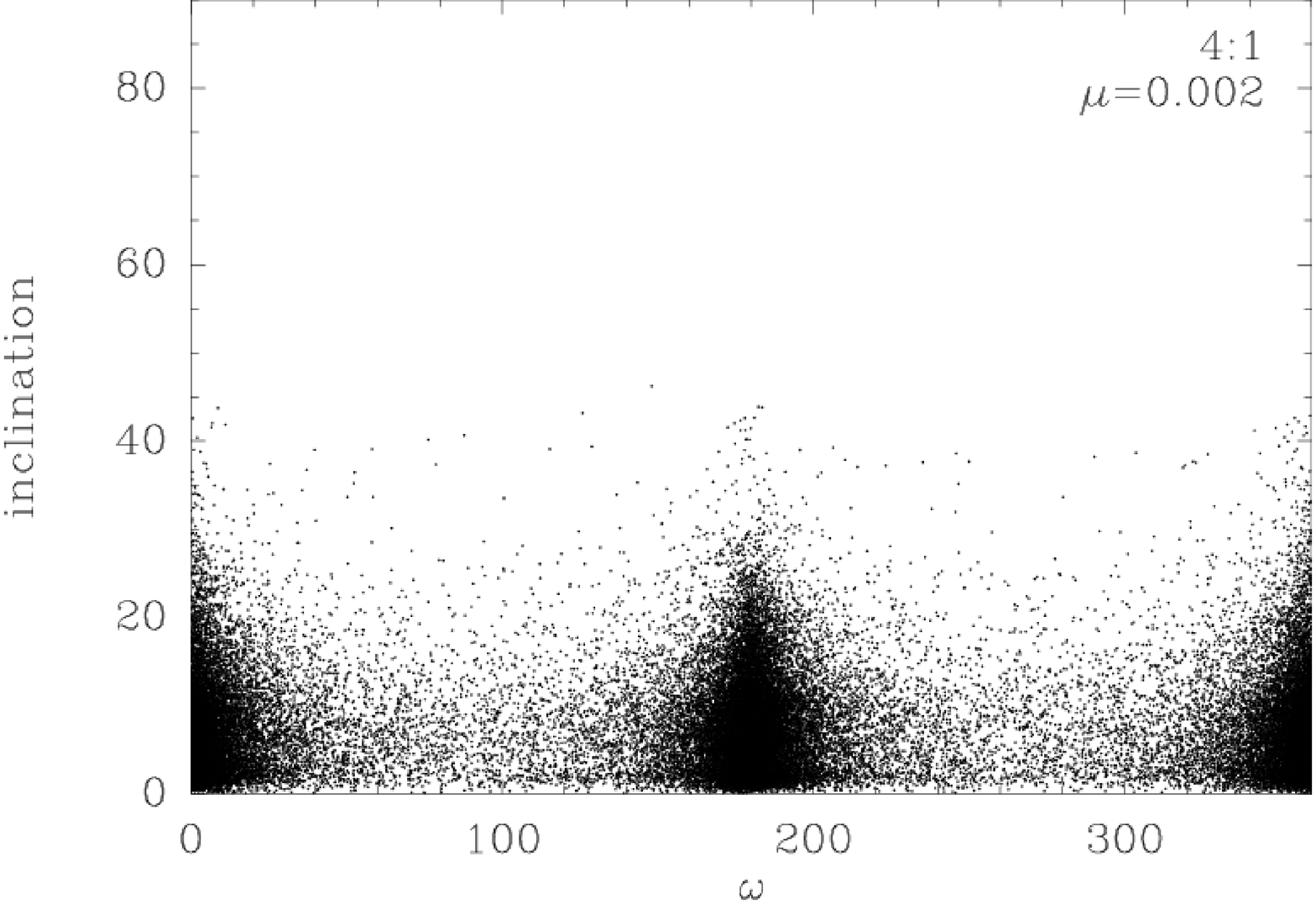}
\caption[]{A statistical test of the inclination oscillation regime for
FEBs. These results concern a typical simulation described in \citet{bm00}
with $e'=0.07$ and $\mu=0.002$. Each dot corresponds to a series of
periastron passages of a body that has entered the FEB regime
(periastron $\la 0.4$), and that is not yet fully evaporated. \textbf{Top
plot :} Inclination as a function of periastron; \textbf{Bottom plot :}
inclination as a function of the argument of periastron $\omega$}
\label{stat}
\end{figure}
To test the statistical effect of the eccentricity jumps
reported, a numerical test over a large number of bodies is necessary.
In fact it was not necessary to perform new simulations. We just
take the (still available) results of the simulations described
in \citet{bm00} and \citet{pth01}. This is illustrated in Fig.~\ref{stat}.
In this figure, we consider a typical simulation described in \citet{bm00},
with $e'=0.07$ and $\mu=0.002$. As described in that paper, the simulation
is made over $10^4$ particles initially chosen orbiting the star in
4:1 mean motion resonances with the perturbing planets. The initial
eccentricities are randomly chosen $\le 0.1$ and the inclinations
$\le 5\degr$. In \citet{bm00}, it was shown that many of the
particles evolve to the FEB regime, and we perform statistics over the
expected Doppler velocities when the FEBs cross the line of sight.
This statistic appears in agreement with the observational one.
In Fig.~\ref{stat}, we display information about the inclinations
of the particles when they are in the FEB regime. More specifically,
we count all periastron passages for the particles with periastron values
less than 0.4\,AU (the FEB regime), and that are not yet destroyed
by evaporation. 
Figure~\ref{stat} shows that the highest inclinations correspond
to the smallest periastron values. This is in agreement with the 
theory outlined above (Fig.~\ref{hseci}) which shows that the high
inclination oscillations are to be expected in the high eccentricity
regime only, i.e. in the FEB regime. Fig.~\ref{stat} also shows
that whenever they reach high inclinations, the FEBs assume
an argument of periastron close to $\pm\pi/2$. This is in 
agreement with the theory, and demonstrates the reality
of the proposed mechanism. Moreover, the concentration of points
close to $\omega\pm\pi/2$ shows that the FEBs spend more time in
high inclination regimes than in low regimes. This is confirmed
directly by the temporal evolution of the inclination in Fig.~\ref{eccinc}.
\section{Stopping the ions}
We showed in the previous section that, due to their inclination
oscillations, the FEBs constitute a potential source of \caii\ and
\fei\ ions
that may escape far off the disk plane.  These ions -- being blown
away by a strong radiation pressure -- should accelerate mostly freely
and reach the outer parts of the disk with high velocities, up to
$\sim 1000\,$km\,s$^{-1}$.  However, such velocities would be in sharp
contradiction to the observations by \citet{bra04}, where the \caii\
ions are found \emph{at rest} with respect to the star, about 116 AU
away.  This implies that some process is able to efficiently slow down
the ions despite the intense radiation pressure they undergo
($\beta=35$, see above).

This issue was investigated recently by \citet{fer06}, as many
metallic species are observed at rest relative to the star despite a strong
radiation pressure. They identified three possible braking processes~:
collisions among ions, collisions with charged ions, and collisions
with a neutral gas. Collisions among ions are very efficient
\citep{ix}, and the whole plasma tends to behave like a single
fluid with a weighted average $\beta$. However, \citet{fer06} show
that unless carbon is overabundant, the fluid is still accelerated
by the star. Collisions with charged grains are only efficient
if the grains are mostly carbonaceous. Moreover, at high latitude
in the disk where the \caii\ is observed, the density of the dust
is low \citep[0.3\%\ of that of the midplane according to the profile
given by][]{fer06}. Collisions with neutral gas are conversely
a good candidate. \citet{fer06} showed that a minimum mass of neutral gas
of $\sim 0.03\,\mbox{M}_\oplus$ is enough to stop the ions.
However, due to the high incoming velocity of the \caii\ ions,
the basic analytic model must be revised. We detail this below, 
and conclude that the actual braking is even more effective than in
the basic formulation.
\subsection{Stopping with gas}
Neutral species like \hi\ or \hei\ or even \hh\ are not subject to any
significant radiation pressure from the star \citep{xxiv}; they may
thus stay orbiting in a Keplerian way at some distance from the star.
The incoming \caii\ ions may then collide into this buffer gas and be
slowed down to negligible velocity.  This process was also invoked as
a way to stop \caii\ ions a few AU from the star in order to generate
the stable circumstellar \caii\ absorption \citep{xxiv}. In that work,
we showed that a column density of $\sim 10^{18}\,\mbox{cm}^{-2}$ is
enough to slow down the \caii\ ions. Here we investigate whether the
same mechanism could also apply to decelerate faster off-plane ions.

The rapid off-plane ions are
not stopped at a few AU like those that stay within the plane
because they do not encounter any noticeable gaseous
medium at their orbital inclination. Why should they be stopped 
around 116 AU ? We must assume that at such a distance, the disk tends
to flare.
Hence some dilute material could be
present at $30\degr$ or more in the outer disk while remaining
absent in the inner disk.  But why 116 AU ? This distance corresponds
approximately
to the location of the power law break-up in the surface brightness
radial profile of the disk \citep{hea00}. Closer to this threshold,
the surface brightness decreases as $r^{-1.1}$, while further away
it falls off much more steeply as $r^{-5.5}$. This was interpreted by
\citet{aug01} as a consequence of the distribution of planetesimals
in the disk. The dust particles are produced by the planetesimals and
then scattered into the outer disk by the stellar radiation pressure.
The power law break-up at $\sim 120\,$AU is consistent with a planetesimal
disk presenting a rather sharp outer edge located at 
this distance \citep{aug01}. The planetesimals disk appears thus to be
truncated at the same distance where the off-plane \caii\ ions stop. 
These facts may be related. The flaring of the disk that we invoke
at that distance for stopping the \caii\ ions could be due to the
perturbations by successive stellar
flybys, as was invoked by \citet{lar01} as an explanation
for asymmetries and arc-like structures in the outer parts of
the circumstellar dust disk. But the same mechanism could also be
invoked to account for the truncation of the planetesimal disk
at the same distance. Further away than 120\,AU, the planetesimals
are perturbed by stellar flybys, and may not remain in a thin
disk. Also the planetesimals themselves may not have had
the opportunity to form there. The stellar flybys may 
have scattered away (and probably in the vertical direction)
the initial material from which the planetesimals were expected
to form. Thus, the \bp\ disk beyond 120\,AU could still
be in a kind of primordial state where no refractory material
condensation would have occurred, with a significant flaring due
to stellar flybys.  
\subsubsection{The analytical induced dipole model and its limitations}
We now review the basic mechanism of decelerating by a neutral medium,
noting that we depart from the situation described in \citet{xxiv}, in
two points: i) the volume density of the incoming ions and of the
colliding medium is probably much less at 116\,AU and $30\degr$
inclination than in the plane at a few AU; therefore no significant
pressure effect is to be expected and the hydrodynamic description may
be dropped; ii) the incoming velocity of the \caii\ ions is likely to
be much higher. In the limiting case of a free ballistic runaway
driven by the radiative pressure of the unobscured star, the resulting
velocity is $\sim 1000\,$km\,s$^{-1}$, about two orders of magnitude
higher than was considered in \citet{xxiv}.

A simple collisional decelerating mechanism was initially described in
\citet{ix} for moderate velocities. When a charged ion approaches a
neutral atom, a dipole is induced on the neutral atom, from which an
interaction results between the two particles that may be well
described by the potential energy
\begin{equation}
V(r)=\frac{1}{4\pi\epsilon_0}\frac{\alpha q^2}{2r^4}\quad,
\label{vdipole}
\end{equation}
where $\alpha$ is the polarizability of the atom, $r$ is the
interaction distance, and $q$ is the charge of the ion, the other
symbols having their usual meaning \citep{mcd64}. The relative motion
in this potential cannot be solved exactly, but there is a critical
impact parameter $b_0$ (that depends on the impact velocity $v$)
separating two regimes~: for $b>b_0$ there is a minimum approach
distance between the particles preventing any close collision; for
$b<b_0$ this is not so, and the particles undergo a physical
collision (see appendix~B). A fairly correct approximation is then to
neglect the effect of the encounter for $b>b_0$, and to consider that
the mean impulsion loss by the ion in the physical collision is $\mu v$,
where $\mu$ is the reduced mass. The interaction cross section is
then $\pi b_0^2$ and the resulting drag force $\vec{f}$ is opposed to
velocity~:
\begin{equation}
\vec{f}=-\mu\pi b_0^2nv\,\vec{v}\qquad,
\label{fcros}
\end{equation}
where $n$ is the number volume density of neutrals. The expression of
$b_0$ (see appendix~B) is
\begin{equation}
b_0=\left(\frac{1}{4\pi\epsilon_0}\frac{4\alpha q^2}{\mu v^2}\right)^{1/4}
\label{b0}
\end{equation}
so that the resulting force is proportional to the velocity:
\mbox{$\vec{f}=-k\vec{v}$} with
\begin{equation}
\label{eq10}
k=n\pi\sqrt{\frac{4\mu\alpha q^2}{4\pi\epsilon_0}}\qquad.
\end{equation}
\begin{table}
\caption[]{Values of the critical impact parameter $b_0$ as a function
of the relative velocity $v$, as computed
from Eq.~(\ref{b0}) for the \caii\ -- \hi\ interaction}
\label{b0tab}
\begin{tabular*}{\columnwidth}{@{\excs}lllll}
\hline
$v$ (km\,s$^{-1}$) & 1 & 10 & 100 & 1000\\
$b_0$ (nm) & 0.785 & 0.24 & 0.078 & 0.024\\
\hline
\end{tabular*}
\end{table}
We confirmed to within 5\% this simple analytical approximation of the
drag force using the more general numerical treatment described later
on.

However this induced dipole regime holds as long as the drift velocity
$v$ is not too large. When $v$ grows, $b_0$ becomes smaller than the
physical size of the particles. In this regime the interaction cannot
be described any longer as attractive, and it rather approaches a hard
sphere regime at shorter range with a constant cross section that does
not depend on $v$. According to Eq.~(\ref{fcros}), the drag force
turns out now to be proportional to $v^2$ instead of
$v$. Table~\ref{b0tab} lists computed values of $b_0$ for different
values of $v$ for the \caii\ -- \hi\ interaction (the polarisability
of \hi\ is $6.7\times10^{-31}$\,m$^3$). We note that as soon as $v\ga
10\,$km\,s$^{-1}$, this induced dipole model becomes unrealistic
because $b_0$ becomes comparable to or smaller than typical atomic radii.
\begin{figure}
\includegraphics[angle=-90,origin=br,width=\columnwidth]{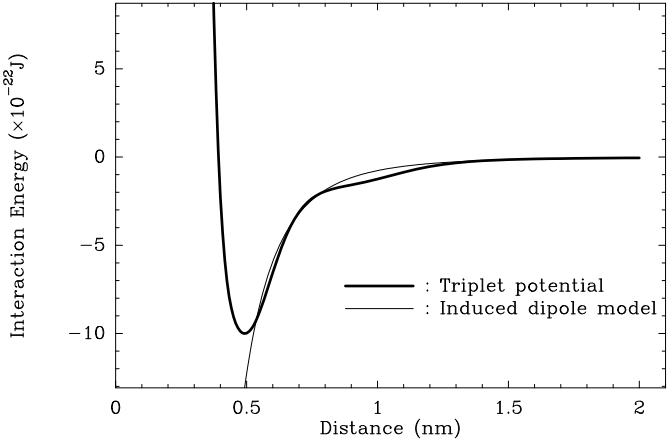}
\caption[]{The ab initio calculated interaction potential between
\caii\ and \hi\ (in their ground triplet state) 
as a function of the mutual distance (fat line),
superimposed on the induced dipole potential (thin line)}
\label{pot}
\end{figure}
Hence \caii\ ions that encounter \hi\ atoms at $\sim
1000\,$km\,s$^{-1}$ undergo a collisional interaction that is similar
to a hard sphere regime. As the velocity decreases 
along successive collisions, the interaction
finally enters the induced dipole regime. A correct description of the
decelerating process of the \caii\ ions implies therefore to be able
to describe the interaction at \emph{every} velocity, in particular in
the intermediate velocity regime between the two above described
extremes. 

\subsubsection{The ``smooth sphere'' model}

In order to have a more coherent description, we 
introduced a ``smooth'' sphere approximation based upon a
continuous description of the interaction potential $U(r)$ between
\caii\ and \hi\ as a function of the relative distance $r$.
The interaction originates from a quantum-mechanical interaction at
the microscopic level.  During a collision, the incoming \caii\ and
\hi\ particles form an intermediate molecular ion. This molecular
complex possess two valence electrons originating from each incoming
particle.  In a quantum description of the interaction, these two
electronic spins recouple to form either a singlet or triplet
state with a 3 to 1 probability in favour of the triplet state. 
However in the case of energetic collisions these input states
will interact with various excited states of the same spin
multiplicity, leading to numerous inelastic processes that will be sketched 
in the the next subsection.

Fortunately the ground state triplet state is not expected to be very
reactive for moderate collisional energies and could provide a
realistic basis for our ``smooth'' sphere model.  Using the {\em ab-initio} 
Gaussian 94 package \citep{g94} we investigated
the triplet input state using a restricted open-shell approach (in
order to preserve the total spin). We performed ROMP2 calculations to
take into account the electronic correlation and also to some extent
the interactions with excited states. We added diffuse and
polarization functions on both centers, paying special attention to
the proper description of the polarisation of the hydrogen atom.  We also
performed a population analysis at the Mulliken level to check that
charge transfer effects were small (contrary to the
singlet state where strong inverse charge transfer effects were
found).  The resulting triplet potential is plotted in Fig.~\ref{pot}
as a function of $r$, superimposed on the induced dipole
interaction. Our ab-initio potential is consistent with the induced
dipole approximation beyond 0.5\,nm, but closer to 0.5\,nm, 
it exhibits a repulsive wall providing a smooth
transition towards the limiting hard sphere regime.

Let us now take into account this ``smooth'' sphere potential for a 
determination of the drag force. 
Given any interaction potential $U(r)$, the drag force acting on the
ion is opposed to the velocity and may be written as
\begin{equation}
\vec{f}=-4\pi n\mu v\left[\int_0^{+\infty}b\,\sin^2\left(
\frac{\chi(b,v)}{2}\right)\,\rd b\right]\,\vec{v}\qquad,
\end{equation}
where $b$ is the impact parameter, and $\chi(b,v)$ is the deflection angle
due to the encounter corresponding to $b$ and $v$. $\chi(b,v)$
itself depends on the form of the potential and may be obtained
from an integral along the relative motion during the encounter
(see appendix~B).
\subsubsection{Beyond the ``smooth sphere'' model: the ``inelastic'' model}
The above expression for the drag force assumes implicitly that the
encounters between the ion and the hydrogen atoms are elastic.

On the contrary, energetic collisions are likely to trigger various
inelastic processes.  For impact velocities in the range
$100$--$1000\,\mbox{km\,s}^{-1}$, the energy available in the center
of mass is huge, from $50\,$eV to $5000\,$eV, and is able to induce a
large variety of excitations in both the valence and core electronic
space. Beyond the mere electronic excitation of either particles,
these energetic collisions can thus trigger a whole range of inelastic
and reactive processes including inverse charge transfer, single or
multiple electronic ionizations, etc\ldots Collisions with \hei\ or
\hh\ will be even more energetic due to the larger reduced mass.  A
detailed theoretical description of all these processes and of their
cross sections and branching ratios is beyond the range of the present
study. Experimental investigations might provide a better starting
point for further studies.

All these inelastic processes will convert a part of the incoming
kinetic energy into internal energy of the particles and also, if
ionization occurs, into kinetic energy of the secondary electrons.  Of
course the internal excitations will mostly decay radiatively and will
never be converted back to kinetic energy of the \caii\ ions.  In
addition, the \caii\ ions are likely to increase their charge,
following either simple or multiple ionization, or inverse charge
transfer with \hi\
\footnote{In view of our ab-initio investigations, this latter process
is expected already to be important at moderate energies for
collisions in a singlet electronic state.}.  Once ionized to \caiii\
or higher, the calcium ions are expected to recombine to \caii\ after
a while. But during the time they spend at higher ionization states,
they should be decelerated even more effectively, because they no
longer feel any significant radiation pressure from the star (as the
species under consideration have no strong spectral lines in the
visible-UV domain), while the drag force is expected to increase with
ionization level.

Moreover, the neutral \hi\ gas, once shocked by the incoming high
velocity ions, is expected to be partly ionized by this process. The
gas will thus tend to behave like a plasma with a collective dynamical
behaviour, resulting in an averaged $\beta$ ratio, as described by
\citet{fer06}. These collective effects should enhance significantly
the efficiency of the braking process.

All these inelastic effects will increase the energy loss by the
\caii\ ions and thus the drag force predicted by the smooth sphere
model. Their description is beyond the scope of the present paper.
 In the following we propose an order-of-magnitude calculation
using a very crude model to describe the ionization processes
involving collisions between \caii\ ions and \hi\ atoms.  A simple way
to treat possible ionization is to monitor the available kinetic
energy $1/2\mu v^2$ before the collision in the inertial referential
frame. First we select close collisions for which the interaction departs 
from the induced-dipole model.  Second we consider the opening of
successive ionization channels when the energy is augmented.  We 
model the ionization for \hi\ and up to 7 electrons for \caii,
assuming an average energy loss $I_\mathrm{p}=20\,$eV per electron.
This loss is supposed to take into account both the extraction 
energy and the kinetic energy of the expelled electron. The ionization
limit of 7 electrons for \caii\ is rather arbitrary and includes the
$2p$ shell and the outer $3s$ electron.  The most energetic collisions
might rather ionize an inner $1s$ electron, but this would result in a
comparable energy loss because their binding energy is higher, about
$150\,$eV.

In practice, we modify the smooth sphere model with 
the following prescription. If
the available kinetic energy exceeds $k\times I_\mathrm{p}$ with
$k\leq 8$ and if the closest approach between the two particles is
less than 0.5\,nm, the incoming 
kinetic energy is arbitrarily reduced by $k\times
I_\mathrm{p}$. This causes the relative velocity after the encounter
$v'$ to be less than the initial velocity $v$. We define the energy
restitution coefficient $e<1$ as $v'=ev$. This may be written as:
\begin{equation}
\frac{1}{2}mv'^2=\frac{1}{2}e^2mv^2=\frac{1}{2}mv^2-kI_\mathrm{p}\qquad.
\end{equation}
If we assume for simplicity that the deflection angle $\chi$ is 
unchanged with respect to the elastic case, then the expression of
the force becomes now
\begin{equation}
\vec{f}=-2\pi n\mu v\left[\int_0^{+\infty}\left(2e\sin^2\left(
\frac{\chi(b,v)}{2}\right)+1-e\right)b\,\rd b\right]\,\vec{v}\qquad.
\end{equation}
Here the coefficient $e$ is implicitly a function of $b$ and $v$, as
explained above.
\begin{figure}
\includegraphics[angle=-90,origin=br,width=\columnwidth]{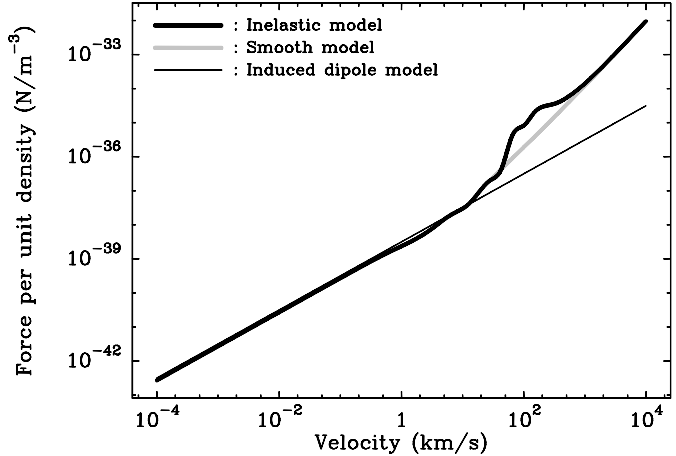}
\caption[]{The drag force on \caii\ ions due to a \hi\ medium of 
unit density ($1\,\mbox{m}^{-3}$), as a function of the drift velocity
$v$. The thin line corresponds to the induced dipole model,
the thick grey line to the smooth model (based upon the triplet potential
shown in Fig.~\ref{pot}) and 
the thick black line to the inelastic model 
(based upon the combination of the triplet potential
and a crude treatment for ionization effects, see text). }
\label{force}
\end{figure}

The result of the force computation in the various cases is shown in
Fig.~\ref{force} as a function of the relative velocity $v$. As in any
case the drag force is proportional to the density of the \hi\ medium
encountered, showing the force for a unit density medium is enough for
comparison purposes. In the induced dipole approximation case, we find
as expected $f\propto v$, and we see that this approximation is valid
up to $v\simeq 20\,\mbox{km\,s}^{-1}$.  At the higher velocity regime,
we have $f\propto v^2$ when the triplet potential is taken into
account, corresponding to our smooth sphere regime.  When ionization
is also taken into account, the non-elastic character of the
interactions adds an extra force term to the elastic smooth sphere
case.  Inelastic effects turns out to be particularly noticeable for
$100\,\mbox{km\,s}^{-1}\la v\la 1000\,\mbox{km\,s}^{-1}$ (at higher
velocity ionization is present but the smooth sphere regime
dominates). This velocity regime concerns
\caii\ ions encountering a \hi\ medium at about 116\,AU.
Consequently, this order-of-magnitude calculation suggests that a
proper inclusion of ionization and other inelastic effects would
significantly enhance the effective drag force beyond the predictions
of the smooth sphere model.
\subsubsection{Estimate of the required \hi\ column density}
Now we have an estimate of the drag force, it is of interest to derive
in which conditions the \hi\ medium is able to stop the \caii\
ions. Irrespective of the initial velocity, if the medium is dense
enough, the ions will always be stopped. The question is to know the
amount of \hi\ neutrals required to do this, and more specifically the
column density $N_\mathrm{s}$ the ions need to cross before being
stopped. Let us first consider a simplified case where radiation
pressure is not taken into account. The ions arrive at initial
velocity $v_0$ and encounter an \hi\ medium with volume density $n$.
The equation of motion of an ion along its path will be
\begin{equation}
m\,\frac{\rd v}{\rd t}=-nF(v)\qquad,
\end{equation}
where $m$ is the mass of the ion, and $F(v)$ is the force per unit density
shown in Fig.~\ref{force}. We change the dependent variable $t$ to the
integrated encountered column density $N$ ($\rd N=nv\,\rd t$):
\begin{equation}
mv\,\frac{\rd v}{\rd N}=-F(v)\qquad,
\end{equation}
from which we immediately derive the integrated column density $N_\mathrm{s}$
required to slow the ions from $v_0$ to $0$:
\begin{equation}
N_\mathrm{s}=m\int_0^{v_0}\frac{v}{F(v)}\,\rd v\qquad.
\label{ns0}
\end{equation}
Let us now reintroduce the radiation pressure as a constant force $P$.
The equation of motion is
\begin{equation}
m\,\frac{\rd v}{\rd t}=P-nF(v)\qquad,
\end{equation}
which is equivalent to
\begin{equation}
mv\,\frac{\rd v}{\rd N}=\frac{P}{n}-F(v)\qquad.
\end{equation}
The ions are now no longer exactly stopped,
but rather slowed down to an equilibrium velocity $v_\mathrm{eq}$
characterised by  $nF(v_\mathrm{eq})=P$. In practice this terminal
velocity is low, so that the ions may be considered as stopped.
The radiation pressure $P$ at 116\,AU on \caii\ ions is
$\sim1.7\times 10^{-29}\,$N (35 times stellar gravity). Let us take
the upper limit of $10^{18}\,\mbox{cm}^{-2}$ for the hydrogen column
density given by \citet{lec01}, spread over a distance $d$. We derive
a ratio
\begin{equation}
\frac{P}{n}=F(v_\mathrm{eq})\simeq 2.5\,10^{-41}\times d(\mbox{AU})
\,\mbox{N}\,\mbox{m}^{-3}\qquad.
\label{veq}
\end{equation}
If we consider as a maximum value for  $d$ a few tens of AU, a comparison
with Fig.~\ref{force} shows that $v_\mathrm{eq}$ is less than
$1\,$km\,s$^{-1}$ and that it falls well within the induced dipole regime.
This result still holds even if we assume a column density lower
by one or even two orders of magnitudes. 

\begin{figure}
\includegraphics[angle=-90,origin=br,width=\columnwidth]{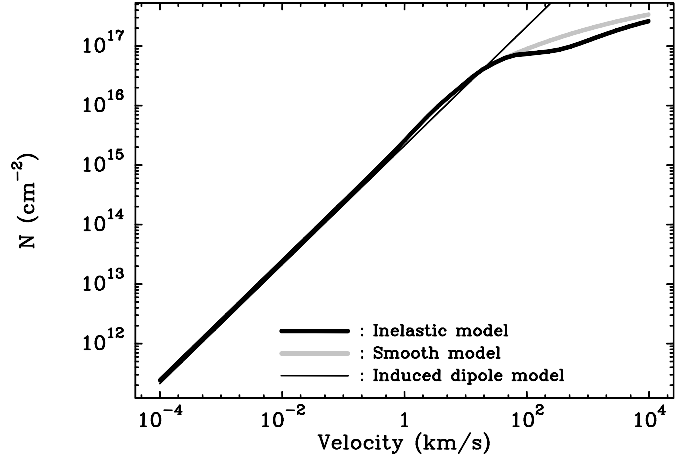}
\caption[]{\hi\ column density necessary to stop \caii\ ions, as a function
of the initial velocity $v_0$, as derived from Eq.~\ref{ns0}, according to 
several models (see text).
The plotting conventions are the same as in Fig.~\ref{force}.}
\label{column}
\end{figure}
Strictly speaking, an infinite column density is required to reach
$v_\mathrm{eq}$. We should write
\begin{equation}
N_\mathrm{s}=m\int_{v_\mathrm{eq}}^{v_0}\frac{v}{F(v)-F(v_\mathrm{eq})}
\,\rd v\qquad.
\label{nsp}
\end{equation}
As close to $v_\mathrm{eq}$, we have $F(v)\propto v$, we derive that
the integral diverges logarithmically towards $v_\mathrm{eq}$.  But
reaching a velocity that is of the same order of magnitude as
$v_\mathrm{eq}$ is enough for our purpose. Moreover, for
$v>>v_\mathrm{eq}$ we have $F(v)>>F(v_\mathrm{eq})$, so that
$F(v_\mathrm{eq})$ can be neglected in Eq.~\ref{nsp}. Finally,
Eq.~(\ref{ns0}) turns out to be a good estimate for $N_\mathrm{s}$
even in the presence of radiation pressure. This is due to the fact
that the terminal velocity $v_\mathrm{eq}$ is very low.

$N_\mathrm{s}$ as given from Eq.~(\ref{ns0}) is plotted on Fig.~\ref{column}
as a function of the initial velocity, for the various interaction
models considered. As expected, in the induced dipole regime,
we have $N_\mathrm{s}\propto v_0$, but at higher velocity, $N_\mathrm{s}$
is reduced by several orders of magnitude with respect to that crude
estimate. 
The smooth sphere model causes $N_\mathrm{s}$ to
stay below a few $10^{17}\,$cm$^{-2}$ (asymptotically 
$N_\mathrm{s}\propto \ln v_0$).
With $v_0=1000\,$km\,s$^{-1}$, we predict
$N_\mathrm{s}\simeq 10^{17}\,\mbox{cm}^{-2}$. This is one order of magnitude
below the upper limit to the \hh\ column density towards \bp\ \citep{lec01}.  
As suggested by our inelastic model, the inclusion of inelastic effects 
would further lower the required column density.  Moreover, the
collective effects decribed by \citet{fer06}, due to partial
ionization of the neutral gas, are expected to enhance the braking
process. $N_\mathrm{s}$ could thus be even less than the
value we derive.

Hence we stress that the model we present here 
provides a plausible mechanism for stopping the \caii\ ions at 100\,AU
from \bp, in order to render them detectable in emission.
\subsection{Stopping with a magnetic field}
Another natural interaction ions that may be subjected to is the influence of
a magnetic field $\vec{B}$ they would encounter at 100\,AU. In the presence
of a magnetic field, the equation of motion of an ion is
\begin{equation}
m\frac{\rd\vec{v}}{\rd t}=\vec{P}+q\,\vec{v}\times\vec{B}\qquad,
\end{equation}
where $\vec{P}$ is the radial radiation pressure. Let us consider for
simplicity that $\vec{P}$ is constant (the ions are supposed to be stopped
over a short distance), as is $\vec{B}$. The equation of motion
has a well known analytical solution. Introducing a referential frame
$(Ox,Oy,Oz)$ where $\vec{B}\parallel Oz$ and $\vec{P}$ lies in the
$xOz$ plane, and the angle $\phi$ between $\vec{P}$ and the $xOy$ plane,
the solution may be described in that referential as
\begin{equation}
\left\{
\begin{array}{rcl}
v_x & = & \omega_g\cos\phi\left[R\cos(\omega_g t)+L\sin(\omega_g t)\right]\\
v_y & = & \omega_g\cos\phi\left[L(\cos(\omega_g t)-1)-R
\sin(\omega_g t)\right]\\
v_z & = & \omega_g\sin\phi\left(L\omega_gt+R\right)\qquad\qquad.
\end{array}\right.
\end{equation}
The ion is supposed to initially move radially at velocity $v_0$.
Here $\omega_g=qB/m$ is the gyromagnetic frequency; $R=mv_0/qB$ is the
gyromagnetic radius associated with the velocity $v_0$; $L$ is defined
as $P=mL\omega_g^2$, i.e., it is a characteristic length associated
with the strength of the radiation pressure.  The motion perpendicular
to the field is a cycloid-like motion, i.e., a combination of a
circular motion of radius $\rho=\sqrt{R^2+L^2}\cos\phi$ at frequency
$\omega_g$ and of a linear drift at velocity $L\omega_g\cos\phi$. The
motion parallel to the field is uniformly accelerated by the radiation
pressure. It is important to note that here, contrary to the neutral
drift model, the velocity does not reach an asymptotic value. If the
field is not perpendicular to $\vec{P}$ ($\sin\phi\neq 0$) the
velocity even increases continuously.  But even if $\sin\phi= 0$ the
motion in the $xOy$ plane is still cycloid-like, and the modulus of
velocity undergoes a periodic modulation of amplitude
$2\rho\omega_g$. Let us now consider a typical magnetic field of
$1\mu G$, at 116\,AU from the star, and
$v_0=1000\,\mbox{km}\,\mbox{s}^{-1}$. We derive
$L=3\,\times10^{-9}\,$AU and $R=0.028\,$AU. Hence $L\ll R$ (i.e., the
Lorentz force dominates the radiation pressure), and
$\rho\simeq R$. This yields 
$2\rho\omega_g\simeq 2v_0=2000\,\mbox{km}\,\mbox{s}^{-1}$.
The ions turn out to have
velocities relative to the star randomly distributed over a range of
$2000\,\mbox{km}\,\mbox{s}^{-1}$.  Even if the ions do not drift away
significantly, their residual velocity range remain far above the
spectral resolution of the instruments, and they should not be
observed as a single line at rest with respect to the star.  In
consequence, this purely magnetic deceleration model cannot account
for the observations.
\subsection{Combining gas drag and magnetic field}
We investigate the possible combination of the two preceding
models, i.e., combining a magnetic field and gas drag. The equation
of motion is now 
\begin{equation}
m\frac{\rd\vec{v}}{\rd t}=\vec{P}+q\,\vec{v}\times\vec{B}+\vec{f}\qquad,
\end{equation}
where $\vec{f}$ is the gas drag force. In the general case where
$\vec{f}$ is given by Fig.~\ref{force}, this equation cannot
be solved analytically, but at low velocity in the induced dipole
regime where \mbox{$\vec{f}=-k\vec{v}$}, the differential system remains
linear and may be solved exactly after a somewhat lengthy but
straightforward algebra. Keeping the definitions and notations
of the preceding sections, the solution is
\begin{eqnarray}
\left\{
\begin{array}{rcl}
v_x & = & \cos\phi\left[(R\omega_g-L\omega_r)\cos(\omega_g t)
+L\omega_g\sin(\omega_g t)\right]\mathrm{e}^{-\omega_rt}\\
&&\qquad\qquad\qquad\qquad\qquad\qquad\qquad\mbox{}
+L\omega_r\cos\phi\\[2\jot]
v_y & = & \cos\phi\left[L\omega_g\cos(\omega_g t)+(L\omega_r-R\omega_g)
\sin(\omega_g t)\right]\mathrm{e}^{-\omega_rt}\\
&&\qquad\qquad\qquad\qquad\qquad\qquad\qquad\mbox{}
-L\omega_g\cos\phi\\[2\jot]
v_z & = & \lefteqn{\dy\sin\phi\,\frac{R\omega_r\omega_g
-L(\omega_r^2+\omega_g^2)}{\omega_r}\,
\mathrm{e}^{-\omega_rt}+L\sin\phi\,\frac{\omega_r^2+\omega_g^2}{\omega_r}
\quad.}
\end{array}\right.
\end{eqnarray}
Here $\omega_g$ and $R$ are defined as above, $\omega_r=k/m$ is a
frequency characterising the drag force; $L$ is now defined as
\mbox{$P=mL(\omega_r^2+\omega_g^2)$}.
The motion is still combination of a circular motion and
a linear drift, but the circular motion is damped exponentially at
frequency $\omega_r$, so that the velocity assumes an asymptotic value
like in the gas drag case without a magnetic field.

Taking now the numbers given above, and considering a typical expected
neutral density of $10^4\,\mbox{cm}^{-3}$ (a column density
of $\sim 10^{17}\,\mbox{cm}^{-2}$ over $\sim 1\,$AU), we derive
$\omega_g\simeq 2.4\times10^{-4}\,\mbox{s}^{-1}$ and
$\omega_g/\omega_r\simeq78$. Hence the exponential damping is a slower
process than the gyromagnetic motion. As $\omega_r\ll\omega_g$,
the numerical value of $L$ is virtually unchanged with respect
to the preceding definition, and we still have $L\ll R$.
After damping,
the ions have a terminal velocity with respect to the gas given by
\begin{eqnarray}
v_\mathrm{t}=\frac{L(\omega_r^2+\omega_g^2)}{\omega_r}
\sqrt{\frac{\omega_r^2+\omega_g^2\sin^2\phi}{\omega_r^2+\omega_g^2}}
& = & \frac{P}{k}\sqrt{\frac{\omega_r^2+\omega_g^2\sin^2\phi}
{\omega_r^2+\omega_g^2}}\nonumber\\
& \simeq & \frac{P}{k}\,\sin\phi\;.
\end{eqnarray}
$P/k$ is the equilibrium velocity $v_\mathrm{eq}$ given by Eq.~(\ref{veq})
in the gas drag case, in the case of the induced dipole
regime. Hence the terminal velocity $v_\mathrm{t}$ appears in any case
less than $v_\mathrm{eq}$. It can even reach zero if the field is 
perpendicular to the initial motion.

Finally, the net result of the interaction is a deceleration of
the ions that is similar to the non-magnetic case. The terminal
velocity is comparable (or even less), and it is reached within
the same characteristic time $1/\omega_r=m/k$. The only difference
concerns the path of the ions. When no magnetic field is present,
the motion of the ions is linear and they are stopped after
having encountered a column density of $10^{17}\,\mbox{cm}^{-2}$,
i.e., $0.7\,$AU with $n=10^4\,\mbox{cm}^{-3}$. With a magnetic field,
the path is no longer linear. When $\phi=0$, the radial extent
of the spiral motion of the ions before being stopped is 
$\simeq 2\rho'$ at $t=0$, i.e., $\simeq 2R$. With the values quoted
above, this is about 0.055\,AU; 
with $n=10^4\,\mbox{cm}^{-3}$, the
corresponding column density is $\sim 8.3\times10^{15}\,\mbox{cm}^{-2}$.
Hence the ions appear to be stopped over a much shorter distance.
This is only due to the fact that the motion is not linear.
As the ions spiral into the quoted distance, they encounter
the required $10^{17}\,\mbox{cm}^{-2}$ to stop them, but not in
a linear fashion, rather inside a box of smaller dimensions.

The magnetic field appears then as an additional source of deceleration
for the ions, but the gas drag remains unavoidable. Our conclusion
is then that a magnetic field may be invoked as a refinement to
the model, but that it is not necessary, the basic process of deceleration
remaining the gas drag.

The main issue concerning the magnetic field should be its sudden
presence around 100\,AU. Why should a magnetic field of $1\,\mu$G appear
there while not present closer to the star ? The only
possibility is to invoke a kind of heliopause. \bpw\ is the only normal
A type star for which a chromospheric activity was detected by
\cite{bou02}. These authors derive a mass loss rate 
\mbox{$\dot{M}=2.5\times 10^{-14}\,M_\odot\,\mbox{yr}^{-1}$}, with a terminal
velocity of $200\,\mbox{km\,s}^{-1}$. Roughly speaking, the suspected
heliopause should be expected where the magnetic pressure $B^2/2\mu_0$
of the surrounding galactic field equals the kinetic pressure
of the wind. With the values of \citet{bou02} and $B=1\,\mu$G, it occurs
at 529\,AU; alternatively, if we want this to occur at 116\,AU, a 
galactic field of $4.5\,\mu$G is required. Inside this cavity, the
field would be radial and have no effect on the motion of the ions.
These are likely values, so that the possibility cannot be excluded.
\section{Conclusion}
The presence of metallic ions at fairly high latitude over the
mid-plane of the \bp\ circumstellar disk, as observed by
\citet{bra04}, can be very well explained as a consequence of the FEB
process. Whenever the evaporating bodies enter the star grazing
regime, they are subject to inclination oscillations up to 
$\sim 30$ -- $40\degr$. The \caii\ ions released by the FEB during this phase
start a free, almost radial expansion pushed by a strong radiation
pressure, keeping track of their initial orbital inclination.  Iron is
also concerned by this process, and we expect \fei\ ions to be present
at high latitude together with \caii. The \fei\ in the data of
\citet{bra04} is not null where \caii\ is detected. We nevertheless
explain in our emission line analysis that, unless the electronic
density is high, the \fei\ emission is expected to be weaker than the
\caii\ one, thanks to a predominant ionization of \fei\ into \feii.

This process does not concern \nai\ ions because, once
produced by the FEBs, they are quickly photoionized into \naii\ and
subsequently no longer experience
any noticeable radiation pressure.
Hence we explain the absence of \nai\ emission at high latitude.

Blown away by strong radiative pressure from the star, the
\caii\ ions reach the distance of $\sim 100$\,AU in about 1\,yr with
final velocities of $\sim 1000\,$km\,s$^{-1}$. They need thus to be
slowed down in order to gather at the star velocity and to form an
observable line. This can be achieved if the ions encounter at that
distance a neutral gaseous medium, in agreement with the conclusions
of \citet{fer06}.
A rough estimate of the incoming ion flux due to
the FEB activity shows that it can account for the necessary heating
source to render the lines observable.

In addition to the induced dipole drag
force considered by \citet{fer06}, we estimated additional braking
effects arising for rapid collisional velocities.  If we consider the
effect of repulsive core and inelastic interactions discussed in Sec
4.1.2 and 4.1.3, a column density of $10^{17}\,\mbox{cm}^{-2}$ of \hi\
is sufficient to stop the ions over a distance of a few AU. The
inelastic model we introduced is probably very crude, but it is still
likely to underestimate the effective drag force.  Therefore
irrespective of the detailed description of the interaction processes,
the required column density remains below the upper detection limit of
$10^{18}\,\mbox{cm}^{-2}$ given by \citet{lec01}.  Following
\citet{fer06}, it should even be less if we took into account the
collective plasma behaviour due to partial ionization of the neutral
gas into account. Conversely, due to the high latitude over the dust
disk, we do not expect collisions with dust grains (invoked
by \citet{fer06} as a possible braking mechanism) to play a significant role
in the decelerating process of the incoming ions.

We also investigate
the possible role of a magnetic field in stopping the ions. While
the sole action of a magnetic field is unable to sufficiently
slow down the ions, magnetic interactions provide an additional
braking process to the basic gas drag model invoked.  Combining gas
drag and magnetic interactions can thus be a very efficient way to
decelerate the ions, still reducing the requirements on the neutral
gas density by an order of magnitude. This nevertheless constitutes a
refinement of the model, as gas drag in itself is sufficient to
account for current observational constraints.

The key parameter in this model is the distance \mbox{($\sim 100$\,AU)}
 at which
the ions are stopped. In the gas drag model, we need to assume that
no neutral medium is present at $30\degr$ inclination
up to that distance, so that the ions can freely expand
radially, and that they suddenly encounter some medium there.
This would mean that the disk begins to significantly flare
at that distance. As explained above, this distance corresponds
also to the expected outer edge of the planetesimal disk that
produces the dust, according to \citet{aug01}. These two facts 
are probably related. 

Our conclusion is thus that the proposed scenario is plausible.
Another important issue in this study
is the number of \caii\ or \fei\ ions necessary to account for the
observations of \citet{bra04}. It cannot be determined easily
even if we may estimate the incoming flux,
as it depends highly on the time the ions stay within the 
neutral medium before diffusing away, and subsequently on
the small asymptotic drift velocity they reach.
If a magnetic field plays a role, this velocity is expected
to be significantly lower than without a field; hence the ions should 
drift more slowly across the neutral medium. At a given epoch,
for the same incoming \caii\ flux
more ions are therefore expected to be trapped in the neutral gas 
if magnetic forces are active than in the opposite case.
This is why deriving an incoming \caii\ flux in order to
compare to the expected number of FEBs is very imprecise. This could
be the purpose of future investigations. Fortunately the uncertainties in
the proposed deceleration models are irrelevant here, because 
once the ions 
have been decelerated, the analytical induced dipole model should be valid.

Our estimate of the incoming ions flux due to FEB activity (Sect.~2) is very
rough, mainly because the FEB activity itself is hard to
constrain. Moreover, we expect this activity to be time-variable, as
changes have been observed between various observing epochs
\citep{tob04}. As the emission lines are supposed to depend on this
flux (via the heating source), we expect the strength of the emission
lines to present temporal variations (at least those at high
latitude). It would thus be of interest to initiate a
follow-up of these lines to check for temporal changes.

The FEB scenario is reinforced by the present analysis.
The off-plane presence of some metallic species, and the absence of
some others, appear as a natural consequence of the FEB scenario
and of the mean-motion resonance model with a giant planet. This
strengthens our view of the \bp\ system as a young planetary system.
%
\appendix
\section{Mean-motion resonance theory}
We describe here the non-planar restricted three-body problem, 
with a mass-less test particle orbiting a star, and perturbed by a
planet orbiting the star. The 
position vectors of the particle and the planet relative to the star
are noted $\vec{r}$ and $\vec{r'}$ respectively. As usual,
we call $\mu$ the ratio of the mass of the planet to the total mass
of the system $m$. We assume that $\mu\ll 1$. In this framework,
the Hamiltonian of the problem is \citep[see, for instance][]{mm93}
\begin{equation}
\mathcal{H}_0
=-\frac{1-\mu}{2a}-\mu\left(\frac{1}{\left|\vec{r}-\vec{r'}\right|}
-\frac{\vec{r}\cdot\vec{r'}}{r'^3}\right)\qquad,
\end{equation}
where $a$ is the osculating semi-major axis of the orbit of the particle.
Here the Hamiltonian has been normalised by the constant factor $\mathcal{G}m$,
where $\mathcal{G}$ is the constant of gravitation.
The usual way is then to start from the classical canonically
conjugate Delaunay variables~:
\begin{equation}
\begin{array}{lcl}
\dy M & , & L=\sqrt{(1-\mu)a}\\[\jot]
\dy \omega & , & G=L\sqrt{1-e^2}\\[\jot]
\dy \Omega & , & H=G\cos i
\end{array}\qquad.
\end{equation}
Here $M$ is the mean anomaly along the orbit of the particle, $\omega$
is its argument of periastron, $\Omega$ its longitude of ascending node,
$e$ its eccentricity and $i$ its inclination with respect to the orbital
plane of the planet. 

We assume here that the particle is locked in a \mbox{$(p+q):p$}
mean-motion
resonance with the planet. It is then of interest to introduce new
canonically conjugate angle-action variables that take into
account the resonance~:
\begin{equation}
\begin{array}{lcl}
\dy \sigma=\frac{p+q}{q}\lambda'-\frac{p}{q}\lambda-\varpi & , &
\dy S=L-G\\[2\jot]
\dy \sigma_z=\frac{p+q}{q}\lambda'-\frac{p}{q}\lambda-\Omega & , & 
\dy S_z=G-H\\[2\jot]
\dy -\nu=\frac{p+q}{q}\lambda'-\frac{p}{q}\lambda & , & 
\dy N=\frac{p+q}{p}L-H
\end{array}
\end{equation}
The modified Hamiltonian is
\begin{equation}
\mathcal{H}=\mathcal{H}_0-\frac{p+q}{p}L\qquad.
\end{equation}
Here $\lambda=M+\omega+\Omega$ is the mean longitude of the particle
along its orbit; $\lambda'$ is the same for the planet;
$\varpi=\omega+\Omega$ is the longitude of periastron. These variables
are introduced by \citet{mm93} and \citet{moo94}. The angle $\sigma$
is often called the ``critical angle of the resonance'' \citep{mm95}.
Resonant orbits are characterised by a 
libration of $\sigma$ around a stable position. 

The secular dynamics is investigated by performing a time-averaging
of $\mathcal{H}$ over the only remaining fast variable, $\lambda'$.
If the orbit of the planet is circular, then the averaged Hamiltonian
turns out to be independent of $\nu$, showing that $N$ is a secular
constant of motion \citep{mm93,mm95}. This can be checked with explicit
expressions, but this arises from the d'Alembert rules~: $\sigma$ and
$\sigma_z$ are independent of any axis rotation within the orbital
plane of the planet, while this is not the case for $\nu$. If the
planet's orbit is circular, the whole Hamiltonian is expected to
be invariant for any rotation in that plane, and should consequently
not depend on $\nu$. In that case, the Hamiltonian
has two degrees of freedom. If we restrict our study to \emph{planar} motion,
then the variables $\sigma_z$ and $S_z$ disappear and the averaged
Hamiltonian is integrable. The secular motion is characterised, together
with the librations of $\sigma$, by coupled oscillations in the $(a,e)$ plane
around the equilibrium value, along a curve $N=\mathrm{cst}$. This dynamics
is described for many specific resonances by \citet{mm93,mm95}.

If the orbit of the planet is not circular, the action $N$ is
no longer constant. It is thus able to evolve, but on a much longer
time-scale than the main librations of $\sigma$. Therefore, on
a short time-scale the oscillations in $(a,e)$ space are preserved,
but on a longer time-scale, the value of $N$ is subject to changes
that may drive the eccentricity to high values. As quoted by
\citet{yosh89}, these changes are particularly important for
resonances 4:1, 3:1 and 5:2. \citet{bm96} showed that
the 4:1 resonance is a potential source of FEBs via this mechanism,
and \citet{pth01} show that the 3:1 resonance may also contribute
to the FEB phenomenon.

If we return to the spatial problem and give an initially moderate
inclination to the particle, these dynamics are preserved. In fact
all the simulations presented in \citet{bm00} and \citet{pth01}
were three-dimensional, and the behaviour reported was in perfect
agreement with the planar description. However, the planar model
does not describe the inclination oscillations observed whenever
the eccentricity reaches high values. In order to explain them,
we must consider the spatial problem as a whole.

To further study the problem, \citet{mm93} introduce
the following canonically conjugate action-angle variables~:
\begin{equation}
\begin{array}{lcl}
\dy \psi_\sigma=\frac{2\pi}{T_\sigma}& , &
\dy J_\sigma=\frac{1}{2\pi}\oint S\,\rd\sigma\\[2\jot]
\dy \psi_z=\sigma_z-\rho_z(\psi_\sigma,J_\sigma,J_z,J_\nu)
& , & J_z=S_z\\[2\jot]
\dy \psi_\nu=\nu-\rho_\nu(\psi_\sigma,J_\sigma,J_z,J_\nu)
& , & J_\nu=N\\
\end{array}\qquad,
\end{equation}
where $T_\sigma$ is the libration period of $\sigma$, $t$ is the time,
and $J_\sigma$ is computed over one libration cycle of $\sigma$.
$\rho_\nu$ and $\rho_z$ are functions that are periodic with zero
average in $\psi_\sigma$. 

We are interested in the \emph{secular} dynamics inside the resonance;
hence we perform a second averaging of $\mathcal{H}$ over $\psi_\sigma$.
The function $\rho_\nu$ and $\rho_z$ disappear then and $J_\sigma$
is a new secular constant of motion. $J_\sigma$ is
the normalised area enclosed by the libration trajectory in $(S,\sigma)$
space. It is close to the amplitude of the libration in $\sigma$.
Thus in the non-circular problem, the value of $N$ may change,
but the amplitude of the libration is roughly preserved. Curves
of $J_\sigma=\mathrm{cst}$ for various resonances are given in
\citet{mm93,mm95}. 

The inclination oscillations are related to the coupled evolution
of $\psi_z$ and $J_z$. This is not easy to describe in the general case
as the Hamiltonian $\mathcal{H}$, even averaged over $\psi_\sigma$,
is still not integrable, as depending on several angles,
and because the relationship between $\psi_z$, $\nu$ and $J_\sigma$ and the
usual orbital elements is not straightforward. A convenient way to
investigate the dynamics is to restrict the study to the case
$J_\sigma=0$, i.e., orbits  with negligible libration amplitude.
In this case, the semi-major axis $a$ assumes a fixed value
(the pericentric equilibrium) close to the unperturbed value
of the resonance; $\sigma$ also is fixed to an equilibrium value
$\sigma_0$. The value of $\sigma_0$ depends on the resonance under
consideration. It is then easy to show that under these conditions
$\psi_z=\sigma_z=\sigma_0+\omega$. This method of considering
zero amplitude libration was used in \citet{bm96}
to draw Hamiltonian maps in the planar problem.

If we consider the circular (but non-planar)
problem, the secular Hamiltonian $\mathcal{H}$, once averaged over
$\psi_\sigma$,  has only one degree
on freedom left. It is a function of $\psi_z$ and $J_z$, or
alternatively of $i$ and $\omega$, the constant value of $N$ 
acting as a parameter. 
\section{The drag force theory}
Given the interaction potential $U(r)$ (the potential energy being $\mu U(r)$),
the encounter between a \caii\ ion and an \hi\ atom may be studied using
energy and momentum conservation~:
\begin{eqnarray}
\frac{1}{2}\left(\dot{r}^2+r\dot{\theta}^2\right)+U(r) & = &
\frac{1}{2}v^2\qquad;\\
r^2\dot{\theta} & = & bv\qquad,
\end{eqnarray}
where the relative motion is described by its polar coordinates
$(r,\theta)$, $v$ is the incoming velocity, $b$ is the impact parameter,
and $\dot{r}$ and $\dot{\theta}$ are as usual shorthands for $\rd r/\rd t$
and $\rd\theta/\rd t$. Eliminating $\dot{\theta}$ leads to the radial equation
\begin{equation}
\dot{r}^2=-2U(r)+v^2\left(1-\frac{b^2}{r^2}\right)\equiv g(r)\qquad. 
\label{eqrm}
\end{equation}
The classical procedure is then to search for a minimum approach distance
$r_\mathrm{m}$. This is done demanding $\dot{r}^2=g(r_\mathrm{m})=0$
in the preceding
equation. There may be no root for $r_\mathrm{m}$. Let us for example
consider the dipole induced potential $U(r)=V(r)/\mu$ where $V(r)$
is taken from Eq.~(\ref{vdipole}). The resulting equation for $r_\mathrm{m}$
is quadratic, and it is a matter of straightforward algebra to show
that a root exists only if $b\ge b_0$, where $b_0$ is given by Eq.~(\ref{b0}).
In the opposite case, the distance is expected to decrease to zero.
However, with a more realistic potential
presenting a repulsive core at small distance, there will \emph{always} be
a root for $r_\mathrm{m}$: at infinity, Eq.~(\ref{eqrm}) gives
$g(r)=v^2>0$ while at small distance, as $U(r)>0$, obviously
Eq.~(\ref{eqrm}) gives $g(r)<0$. 

$r_\mathrm{m}$ corresponds to a polar angle $\theta_\mathrm{m}$, and it
is easy to see that the defection angle $\chi$ is related to
$\theta_\mathrm{m}$ by
\begin{equation}
\chi=2\theta_\mathrm{m}+\pi\qquad.
\end{equation}
$\theta_\mathrm{m}$ itself can be deduced from the fact that at infinity
before the encounter, we have $\theta=-\pi$:
\begin{equation}
\theta_\mathrm{m}+\pi=\int_{r_\mathrm{m}}^{+\infty}
\left|\frac{\rd\theta}{\rd r}\right|\,\rd r=
\int_{r_\mathrm{m}}^{+\infty}
\frac{\dot{\theta}}{|\dot{r}|}\,\rd r
=\int_{r_\mathrm{m}}^{+\infty}
\frac{bv}{r^2}\frac{\rd r}{\sqrt{g(r)}}\;,
\end{equation}
where $g(r)$ is defined in Eq.~(\ref{eqrm}). Setting $y=r_\mathrm{m}/r$,
this expression may be rewritten as
\begin{equation}
\theta_\mathrm{m}+\pi=\frac{b}{r_\mathrm{m}}\int_{0}^{1}
\frac{\rd y}{\sqrt{1-y^2+\frac{2}{v^2}\left[y^2U(r_\mathrm{m})
-U(r_\mathrm{m}/y)\right]}}\qquad,
\end{equation}
from which we derive straightforwardly
\begin{equation}
\chi(b,v)=\frac{2b}{r_\mathrm{m}}\int_{0}^{1}\frac{1}{\sqrt{1-y^2}}
\frac{\rd y}{\sqrt{1+\epsilon(b,v,y)}}-\pi\qquad,
\end{equation}
where
\begin{equation}
\epsilon(b,v,y)=\frac{2}{v^2}\frac{y^2U(r_\mathrm{m})-U(r_\mathrm{m}/y)}
{1-y^2}\qquad.
\end{equation}
This last expression is of practical interest to numerically compute $\chi$
for any potential: $\epsilon(b,v,y)$ has a finite limit towards
$y\rightarrow 1$, so that the presence of the kernel $1/\sqrt{1-y^2}$
allows a rapid computation using a Gauss-Chebychev quadrature technique.

Once $\chi$ is known, the impulsion change to the ion during
the encounter reads 
\begin{equation}
\delta\vec{p}=\mu v\left|\begin{array}{l}\cos\chi-1\\\sin\chi\cos\phi\\
\sin\chi\sin\phi\end{array}\right.\qquad,
\end{equation}
once written in a Cartesian $(x,y,z)$ referential frame with the $x$-axis
parallel to the initial motion of the ion, and where $\phi$ is an
azimuthal angle characterising the plane of the encounter. Of course
over many encounters $\phi$ is a random angle, so that the $y$ and $z$
components of $\delta\vec{p}$ vanish. On average then, we have
\begin{equation}
\delta\vec{p}=\mu (\cos\chi-1)\,\vec{v}=-2\mu\sin^2\left(\frac{\chi}{2}\right)
\,\vec{v}\qquad.
\end{equation}
Then follows a classical cross-section calculation. The number of encounters
occurring within the time-span $\delta t$ at $(b,\phi)$ within
$\rd b$ and $\rd\phi$ is $n v\,\delta t\,b\,\rd b\,\rd\theta$. 
Multiplying then $\delta\vec{p}$ by this number of encounters
and integrating over $\theta$ and $b$, we derive the drag force as:
\begin{eqnarray}
\vec{f} & = & \frac{1}{\delta t}\int\!\!\!\int \delta\vec{p}(b,v)\,n v
\,\delta t\,b\,\rd b\,\rd\theta\nonumber\\
& = & -4\pi n\mu v\left[\int_0^{+\infty}b\,\sin^2\left(
\frac{\chi(b,v)}{2}\right)\,\rd b\right]\,\vec{v}\qquad,
\end{eqnarray}
If the collision is not elastic, the modulus of the relative
velocity $v'$ after the encounter differs from the initial one
$v$. We have $v'=ev$. The impulsion change to the ion during
the encounter reads now
\begin{equation}
\delta\vec{p}=\mu \left|\begin{array}{l}v\cos\chi-ev\\v\sin\chi\cos\phi\\
v\sin\chi\sin\phi\end{array}\right.\qquad,
\end{equation}
As in the elastic case, there is no average change perpendicular to the
motion. In the direction of the motion, we derive
\begin{equation}
\delta\vec{p}=\left(-2\mu e\sin^2\left(\frac{\chi}{2}\right)+(e-1)\mu\right) 
\,\vec{v}\qquad.
\end{equation}
Integrating as above over $b$ and $\theta$ we derive the drag force
as
\begin{equation}
\vec{f}=-2\pi n\mu v\left[\int_0^{+\infty}\left(2e\sin^2\left(
\frac{\chi(b,v)}{2}\right)+1-e\right)b\,\rd b\right]\,\vec{v}\qquad.
\end{equation}.

\begin{acknowledgements}
All the computations presented in this paper were performed at the 
Service Commun de Calcul Intensif de l'Observatoire de Grenoble (SCCI).
Comments by our referee, V. Grinin, and by the Editor, M. Walmsley, 
inspired the emission model presented in Sect~2. 

\end{acknowledgements}

\end{document}